\documentclass[twocolumn,showpacs,preprintnumbers,amsmath,amssymb,superscriptaddress,aps]{revtex4}

\usepackage{graphicx}
\usepackage{dcolumn}
\usepackage{bm}
\usepackage[caption=false]{subfig}
\usepackage{epsfig}
\usepackage{graphics}
\usepackage{mathtext}

\begin{document}

\title{Dynamic Scaling, Data-collapse and Self-Similarity in Mediation-Driven Attachment Networks}

\author{Debasish Sarker}
\affiliation{Department of Physics, University of Dhaka, Dhaka 1000, Bangladesh}
\author{Liana Islam}
\affiliation{Department of Physics, University of Dhaka, Dhaka 1000, Bangladesh}
\affiliation{Department of Biophysics, Johns Hopkins University, Maryland, USA }
\author{Md. Kamrul Hassan} 
\affiliation{Department of Physics, University of Dhaka, Dhaka 1000, Bangladesh}

\begin{abstract}%
Recently, we have shown that if the $i$th node of the Barab\'{a}si-Albert 
(BA) network is characterized by the generalized degree $q_i(t)=k_i(t)t_i^\beta/m$, where $k_i(t)\sim t^\beta$
and $m$ are
its degree at current time $t$ and at birth time $t_i$, then the corresponding distribution function $F(q,t)$ exhibits dynamic scaling. 
Applying the same idea to our recently proposed mediation-driven attachment (MDA) network, we find 
that it too exhibits dynamic scaling but, unlike the BA model, the exponent $\beta$ of the MDA model assumes
a spectrum of value $1/2\leq  \beta \leq 1$. Moreover, we find that the scaling curves for small $m$ are significantly different 
from those of the larger $m$ and the same is true for the
BA networks albeit in a lesser extent. We use the idea of the distribution of inverse harmonic mean (IHM) of the neighbours of 
each node and show that the number of data points that follow the power-law degree distribution
increases as the skewness of the IHM distribution decreases.  
Finally, we show that both MDA and BA models become almost identical for large $m$.
\end{abstract}

 
\maketitle

\section{Introduction}

Many complex systems can be described as an interwoven web of large network if the constituents 
are regarded as nodes or vertices and the interactions between constituents as links or edges. 
For example, the human brain is a network of neurons linked by axons,
cells of living systems are networks of molecules linked by chemical interaction, 
the Internet is a network of routers and computers linked by cables or wireless connections, 
the power-grid is a network of substations linked by transmission lines, the World Wide Web (WWW) 
whose nodes are HTML document connected by URL addresses \cite{ref.protein, ref.internet, ref.www}.
Equally, there are social networks where individuals are nodes or vertices 
linked by social interactions like friendships, professional ties etc \cite{ref.coauthorship, 
ref.movieactor}. The notion of networks was born as graph in 1735 when the Konigsberg's seven bridge
problem was solved by Leonard Eular. It was eventually developed as graph theory and became 
an active subject of discrete mathematics \cite{ref.Bollobas}.
However, the first major breakthrough was made by Paul Erd\"{o}s and Alfred R\'{e}nyi in 1959 
\cite{ref.erdos}. They realized that the real-life network must have some degree of disorder. 
To that end, they assumed that pairs of nodes are picked at random 
from a fixed number of labeled nodes and links are established with some probability $p$. 
Such a simple model proved to have interesting properties and considered to have the potential 
of describing real life networks. The main result of the  Erd\"{o}s-R\'{e}nyi (ER) model 
is that the degree distribution $P(k)$, the probability that a randomly chosen node is connected to $k$ other nodes by one edge,  
is Poissonian. However, real networks are neither completely regular where every node has 
the same degree nor completely random where the degree distribution is Poissonian.
 
A second breakthrough, or rather a paradigm shift
occurred in 1999  exactly $40$ years after the ER model thanks to Barab\'{a}si and Albert. They 
not only revolutionized the notion of the graph theory but also rebranded it under a new name
{\it network} instead of graph. Besides, they realized the fact that natural and man-made networks are 
neither static nor they establish links randomly \cite{ref.barabasi, ref.barabasi_1}. 
Instead, they are formed by continuous addition of new nodes such that the incoming nodes tend
to establish
links with an existing one by choosing it preferentially with respective to their
degree of connectivity.  Such preferential attachment (PA) rule
 essentially embodies the intuitive idea of the {\it rich get richer} 
principle of the Matthew effect in sociology \cite{ref.barabasi}. 
Incorporating these two simple rules, namely growth and PA rule, Barab\'{a}si and Albert (BA) 
then presented a simple theoretical model and showed that the resulting 
network can reproduce the power-law degree distribution which most real life 
networks exhibit. The impact of this seminal article in developing network science
as a truly interdisciplinary  subject can hardly be exaggerated.

Despite the fact that the BA model can capture the generic features of many 
seemingly unrelated real world or man-made networks, it also has 
some shortcomings like most groundbreaking models. First, each new node must 
know the degree of all the nodes of the 
existing network so that it can choose one from these, which is a formidable task especially when
the network size $N\rightarrow \infty$. Second, 
the exponent of the degree distribution assumes a fixed value $\gamma=3$ independent of the
number of links $m$ with which each new node is born while most natural and man-made networks 
have $2<\gamma\leq 3$. Recently, we have proposed an alternative model which can
address both the drawbacks  \cite{ref.hassan_liana}. In this model, each new node comes with $m$ links. Then it first 
picks randomly a mediator from the existing network and connects itself not with the mediator 
but with $m$ of its neighbours, which are 
also picked at random with uniform probability. It has been shown that such mediation-driven 
attachment (MDA) rule is super preferential for small $m$  and the extent of preference of choosing
highly connected ones get weaker as $m$ increases. Despite the
fact that the nodes with higher degree are more likely to gain links with new nodes than those
with lower degree, each of the existing node still follow the same growth law $k_i(t)\propto t^{\beta(m)}$
where the exponent $\beta(m)$ depends on $m$. The growth law of the BA model also has the same form but
the exponent $\beta=1/2$ independent of the value of $m$.
Such growth law is a signature that
the degree distribution exhibits a power-law at least near the tail.

 In 2011 Hassan et al. showed that the generalized degree $q$, which is the product of the degree 
of a given node $k$ and the square root of its birth time, is also an interesting
parameter \cite{ref.hassan_ba_dc}. It can characterize
the nodes of the growing network better than the degree of its node itself. 
It has been shown then that the generalized degree distribution function $F(q,t)$ exhibits
dynamic scaling $F(q,t) \sim t^{-\beta} \phi (q/t^\beta)$ where $\beta=1/2$
and $\phi (x)$ is the scaling function. In this work, we apply the same idea to the MDA model
and compare the results with those of the BA model. 
First, we show that, like in the BA model, the nodes of the MDA network too 
 grow with time $k\sim t^{\beta(m)}$ but unlike the BA model the growth exponent $\beta$ 
assumes values within $1/2\leq \beta < 1$ depending on $m$. 
Thus, the nodes of the MDA model too can be  
characterized by the generalized degree, the product  of their degree and the corresponding birth time
 raised to the power $\beta$. Second, we show that the generalized degree distribution for the
MDA model  also exhibits dynamic scaling like BA model. Third, the MDA network is highly sensitive to 
the value of $m$ as we find that for each value of  $m$ the resulting network belongs to 
different universality classes as both their scaling functions and exponents are distinct and different. 
This is, however, not the case for the BA network since the growth exponent $\beta$ is always the same 
although the nature of the universal scaling curves for the small $m$ values is significantly different
from that for large $m$ values.

The rest of the article is organized as follows. In section II, we give the algorithm
of the model as it can describe the model in a much better way than its mere definition. In the next section, 
we give the mathematical formulation of the model and obtain not only the analytical solution for degree 
distribution but also discusses its various aspects including its second moment. In section IV, we discuss 
dynamic scaling and to prove it we invoke the idea of data-collapse. In section V, the connection between dynamic scaling, data
collapse and self-similarity is discussed citing the example of geometric similarity. Finally, in section VI we
give a summary of the article. 

\section{The model}

The growth of the MDA network starts from a seed of size $m_0$ which consists of a small number of 
nodes connected in an arbitrary fashion. However, the minimum value of $m_0$ has to be 
equal to degree $m$ of the newborn nodes. Once the seed is chosen the 
network then grows according to the following algorithm:

\begin{enumerate}

 \item[i] Choose an already connected existing node at random with uniform probability and regard it as
 the mediator.
 
 \item[ii] Pick $m$ of its neighbors also at random with uniform probability.
 
 \item[iii] Connect the $m$ links of the new node with the $m$ neighbors of the mediator.
 
 \item[iv] Increase time by one unit.
 
 \item[v] Repeat steps (i)-(iv) till the desired network size is achieved.
\end{enumerate}

The basic definition of the MDA rule is not completely new as it has first  been  studied by Yang {\it et al.}
\cite{ref.Yang}. However, their mathematical formulation and results are totally different from our findings in Ref. \cite{ref.hassan_liana}.
It is noteworthy to mention that  we conceived the idea of the MDA while we were working on weighted planar stochastic lattice (WPSL) \cite{ref.hassan_njp, ref.hassan_conf}. 
It has been shown that the dual of WPSL, where
 an existing node gains links only if one of its neighbors is picked, emerges as a network with power-law degree
distribution. We then became curious as to what happens if a graph is grown following a similar
rule. However, it took a long time to find an exact mathematical form for probability that an existing node is finally picked through a mediator to connect with one of the links of the
new node. In the following section we give the mathematical formulation and discuss its various aspects.

\section{Mathematical Formulation}

We first write an expression for the probability $\Pi(i)$ that an arbitrary node $i$ of the existing
network is picked to get connected to one of the links of the incoming node. Say, the node $i$ has 
degree $k_i$ and hence it has $k_i$ mediators. Each of these mediators are picked with probability
$1/N$ and from each of those mediators we can pick the node $i$ with probability equal 
to the inverse of their respective degrees. We can therefore write
\begin{equation} 
\label{eq:1}
\Pi(i)= \frac{\sum_{j=1}^{k_i}{\frac{1}{k_j}}}{N},
\end{equation}
where $\sum_{i=1}^N\sum_{j=1}^{k_i}{\frac{1}{k_j}}=N$ and hence the probability $\Pi(i)$
is naturally normalized. Then the degree $k_i$ of the node $i$ evolves according to the following 
rate equation
\begin{equation} 
\label{eq:2}
{\frac{\partial k_i}{\partial t}} = m \ \Pi(i).
\end{equation}
The factor $m$ takes care of the fact that any of the $m$ links of the newcomer may connect
with the node $i$. Solving Eq. \eqref{eq:2} when $\Pi(i)$ is given by Eq. \eqref{eq:1} seems
quite a formidable task unless we can simplify it.

The spirit of the MDA rule can be found in some of the earlier works as well.
For instance, the work of Saram\"{a}ki and Kaski partially overlaps with our work as a 
special case \cite{ref.mda_2}. However, the expression for the attachment
probability that they obtained and the corresponding results they found are
 totally different from our results.
On the other hand, the spirit of one of the models proposed by Boccaletti {\it et al.} may also
appear similar to ours albeit markedly different on closer look \cite{ref.boccaletti}. In their model
the incoming nodes 
have the option to connect either to the mediator or to one of their neighbors while in our model
the new node can connect only to the neighbours of the mediator. Nevertheless, they too found that the degree distribution 
exhibits power-law with the same exponent $\gamma=3$ as that of the BA model
independent of the value of $m$, which is again far from what our model entails.

Yet another closely related model is the Growing
Network with Redirection (GNR) model proposed by Gabel, Krapivsky and Redner where at each time
step a new node either attaches to a randomly chosen target node with probability $1-r$ or
to the parent of the target with probability $r$ \cite{ref.redirection}. The GNR model with
$r=1$ may appear similar to our model but it should be noted that unlike the GNR
model our MDA model is for undirected networks. One more difference is that, in our model new nodes may
join the existing network with $m$ edges whereas they considered $m=1$ case only. We have already shown
that $m$ plays a very crucial role. Another closely related model also proposed by
 Krapivsky and Redner is known as  growing network with copying (GNC) model \cite{ref.gnc}. 
In this GNC model, a new node attaches to a randomly selected node, as well as to all 
the neighbors of the mediator. This is quite similar to the  model studied by Boccaletti except the
fact that GNC model produces directional network.

\begin{table}[h!]
\centering
    \begin{tabular}{| l | l | l |}
    \hline
    $m$ & $\beta$ for MDA model & $\beta$ for BA model \\ \hline
    $1$ & $0.999229$ & $0.5$  \\ \hline
    $3$ & $0.996710$ & $0.5$  \\ \hline
    $11$ & $0.886872$ & $0.5$  \\ \hline
    $12$ & $0.845995$ & $0.5$  \\ \hline
   $ 13$ & $0.822789$ & $0.5$ \\ \hline
$25$ & $0.630948$ & $0.5$ \\ \hline
$50$ & $0.547211$ & $0.5$ \\ \hline
$100$ & $0.519304$ & $0.5$ \\ \hline
$m>>100$ & $0.5$ & $0.5$ \\

    \hline
    \end{tabular}
\caption{The kinetic exponent $\beta$ of the MDA and BA model for different $m$ values.}
\label{table:1}
\end{table}

\begin{figure}

\centering

\subfloat[]
{
\includegraphics[height=2.4 cm, width=4.0 cm, clip=true]
{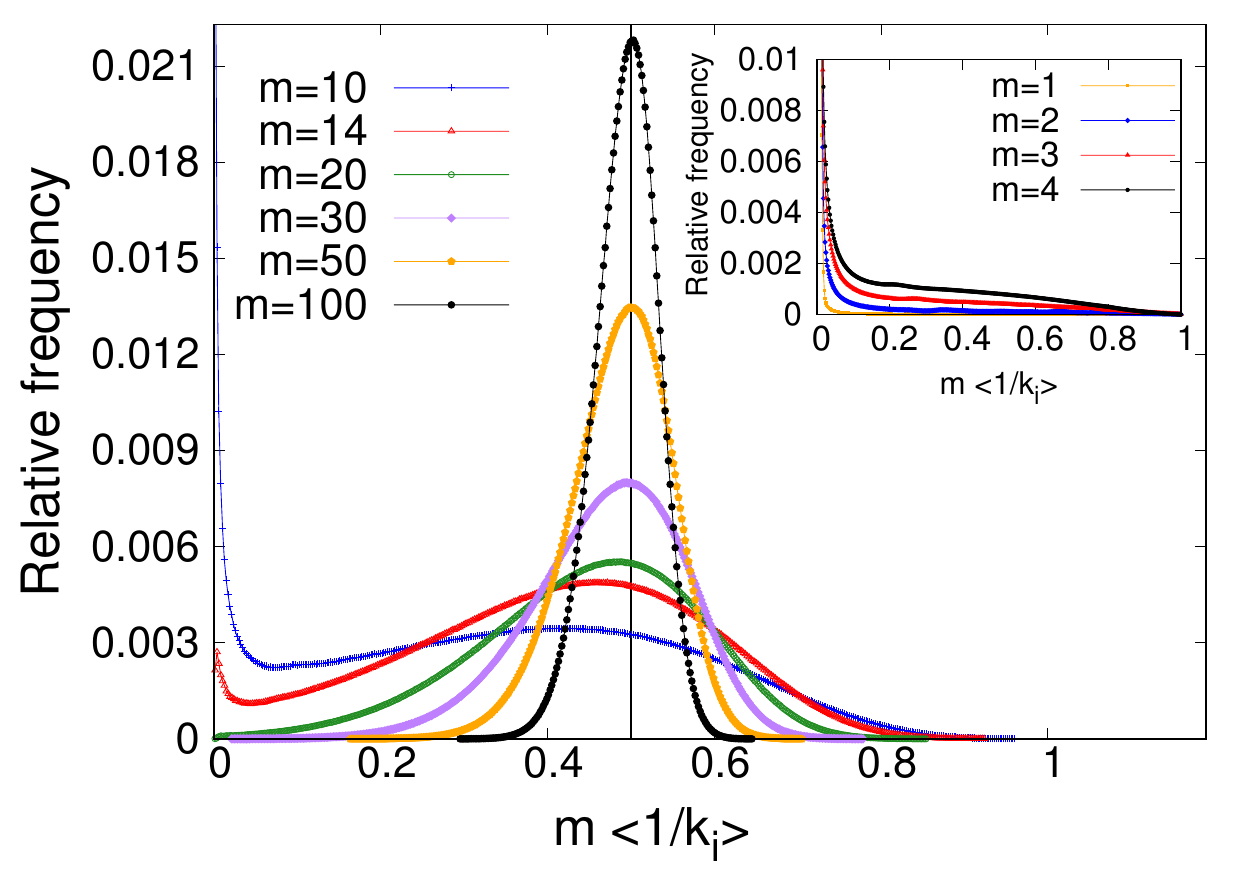}
\label{fig:1a}
}
\subfloat[]
{
\includegraphics[height=2.4 cm, width=4.0 cm, clip=true]
{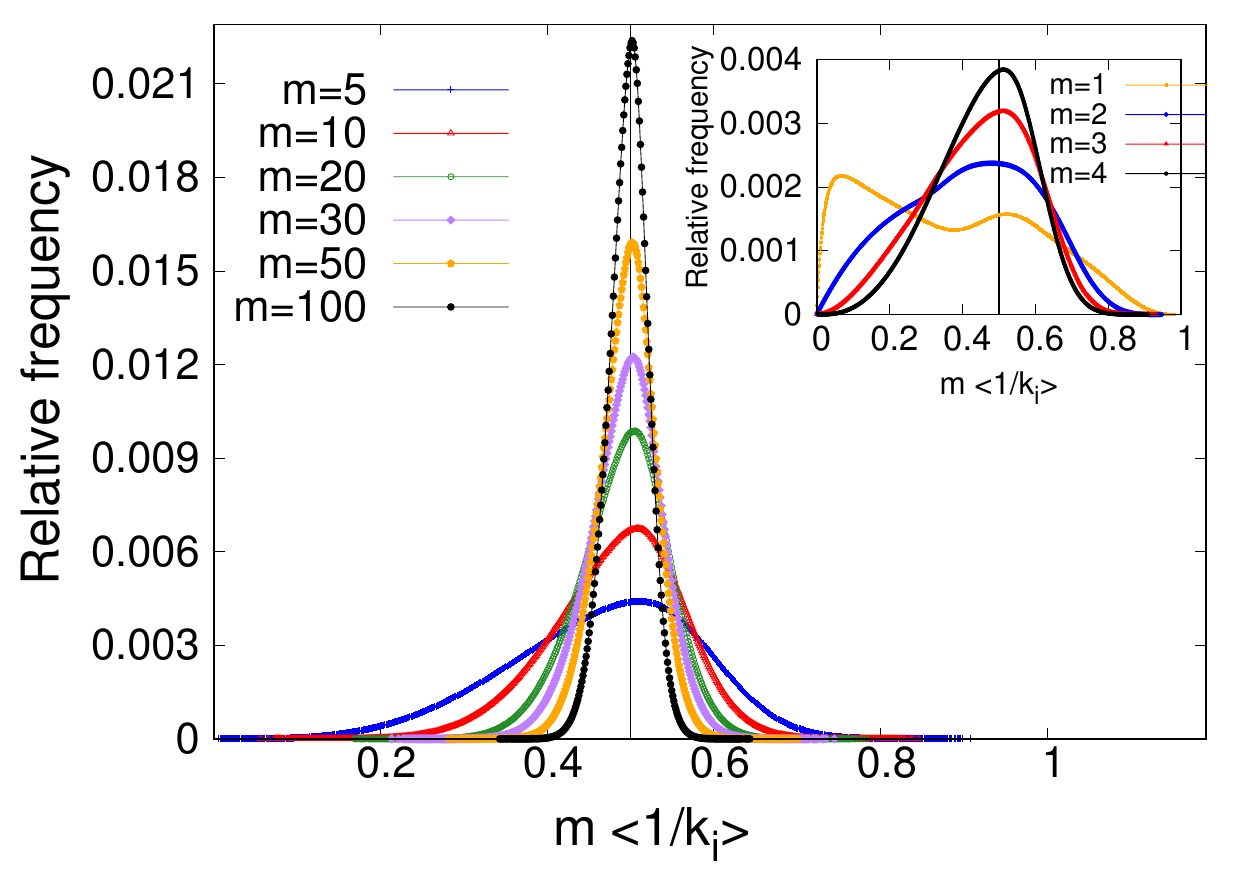}
\label{fig:1b}
}

\caption{
Relative frequency distribution of IHM value for different $m$ are shown in (a) for MDA model
and (b)for BA models. The same plots for small $m$ are shown in the inset of respective plots.
} 

\label{fig:1ab}
\end{figure}

In order to solve Eq. (\ref{eq:2}) we find it convenient to
re-arrange Eq. (\ref{eq:1}) as follows
\begin{equation} \label{eq:npi}
\ \Pi(i) = \frac{k_i}{N} \ \Big (\frac{\sum_{j=1}^{k_i}{\frac{1}{k_j}}}{k_i} \Big ).
\end{equation}
The factor within the bracket is known as the inverse of the harmonic mean
(IHM) of degrees of all the $k_i$ neighbours of the mediator $i$. Now if we can show that
this quantity in some situations can be approximated as a constant then we argue that it embodies the 
intuitive idea of the preferential attachment (PA) rule. To that end, we have performed extensive
numerical simulation to know the nature of IHM as a function of $m$ and $N$.  We
find that for small $m$, specially for $m=1,2, 3$ etc., 
the relative frequency distributions, fraction of the total nodes which have IHM value within a 
given class of the IHM values are totally different from those for higher $m$ values. This is shown
in Figs. (\ref{fig:1a}) and (\ref{fig:1b}) for MDA and BA models respectively. 
We can clearly see from these figures that
for small $m$ the IHM values of the MDA model are not peaked around its mean but the same is not true for the BA model.  In the 
case of the BA model, except for $m=1$, the distribution for all $m$ has at least one peak 
around which its value fluctuates. Moreover, as the $m$ value increases, the peak is increasing more pronounced. 
The product of the value of IHM at which the peak of the relative frequency occurs 
and the corresponding value of $m$, which is the $\beta$, is always equal to $1/2$ for
the BA model. Interestingly, 
this $\beta$ value always coincides with the mean regardless of the value of $m$.
However, the wide variance of this distribution has its signature in the degree distribution. Note that the
minimum condition to have a good power-law degree distribution is to have normal distribution of
relative frequency distribution of the IHM value of all the nodes in the network. Besides, the lesser the
variance the better the power-law degree distribution.

In the case of MDA model, we begin to get bell-shaped curve only when $m>12$ 
but they are left skewed bell shaped curve for up to say $m=30$. That is, 
as $m$ increases beyond $m=12$, the relative frequency distributions
gradually become more symmetric around the maximum value which eventually occurs at the mean
IHM. Note that for small $m$ the peak and the mean do not coincide which is in sharp contrast with the
BA model. Moreover, as $m$ increases, we find that the fluctuations occur at increasingly lesser extents,
and hence the mean IHM value becomes the characteristic value of the entire network. Also interesting is the fact
that for a given $m$ we find that the mean IHM value in the large $N$ limit becomes constant. Using these two 
factors  we can apply the mean-field approximation (MFA). That is,  we can replace the true
IHM value of each node by their mean and hence the attachment probability $\Pi(i)$
becomes linear with degree. Using this in Eq. (\ref{eq:npi}) we can write Eq. (\ref{eq:2}) as
\begin{equation} 
\label{eq:3}
{\frac{\partial k_i}{\partial t}} = k_i \ \frac{\beta (m)}{t}.
\end{equation}
Here we assumed that the mean IHM is equal to $\beta(m)/m$ for large $m$ where the factor $m$ in the
denominator is introduced for future convenience. It is noteworthy to mention 
that the size $N$ of the network is an indicative of time $t$ since we assume
that only one node joins the network at each time step. Thus, for $N>>m_0$ we can write
$N \sim t$.

\begin{figure}

\centering

\subfloat[]
{
\includegraphics[height=2.4 cm, width=4.0 cm, clip=true]
{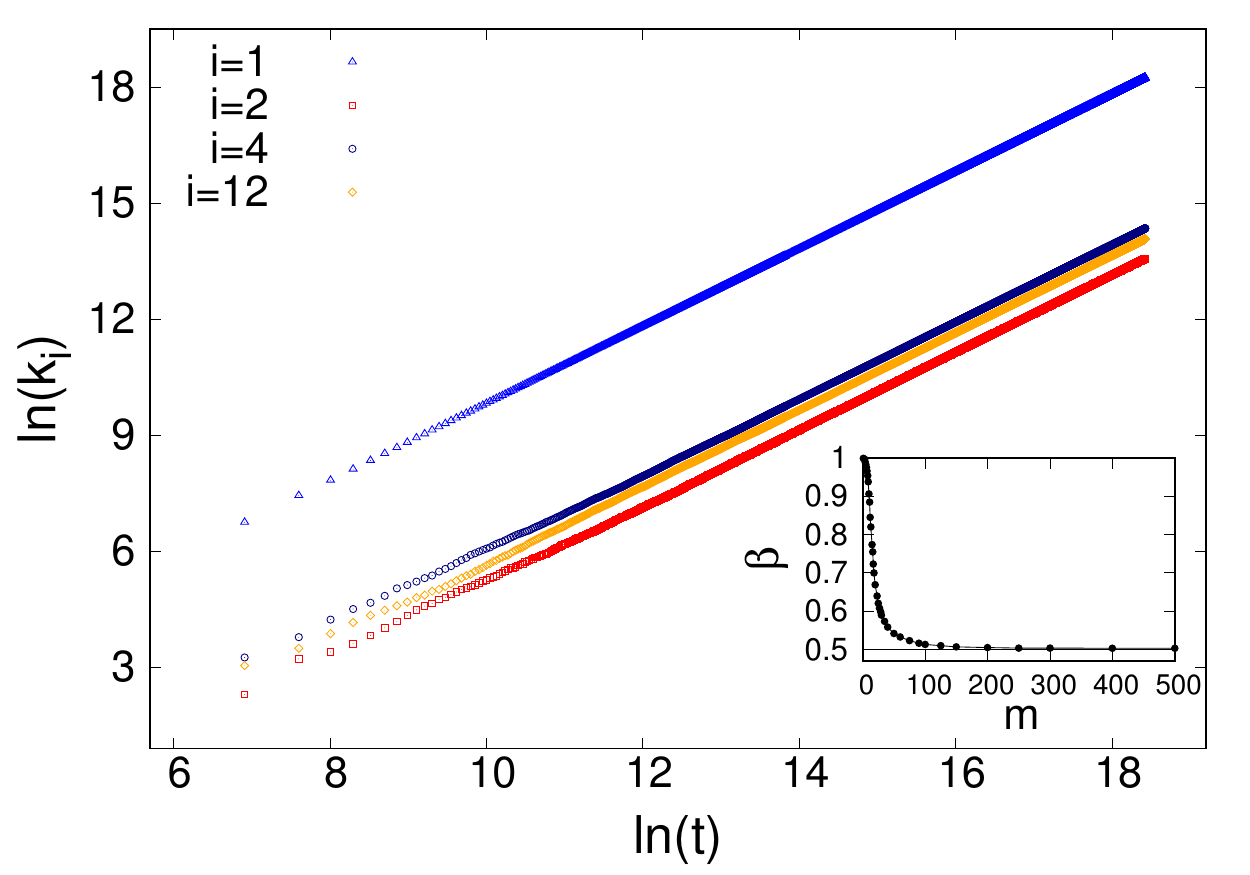}
\label{fig:2a}
}
\subfloat[]
{
\includegraphics[height=2.4 cm, width=4.0 cm, clip=true]
{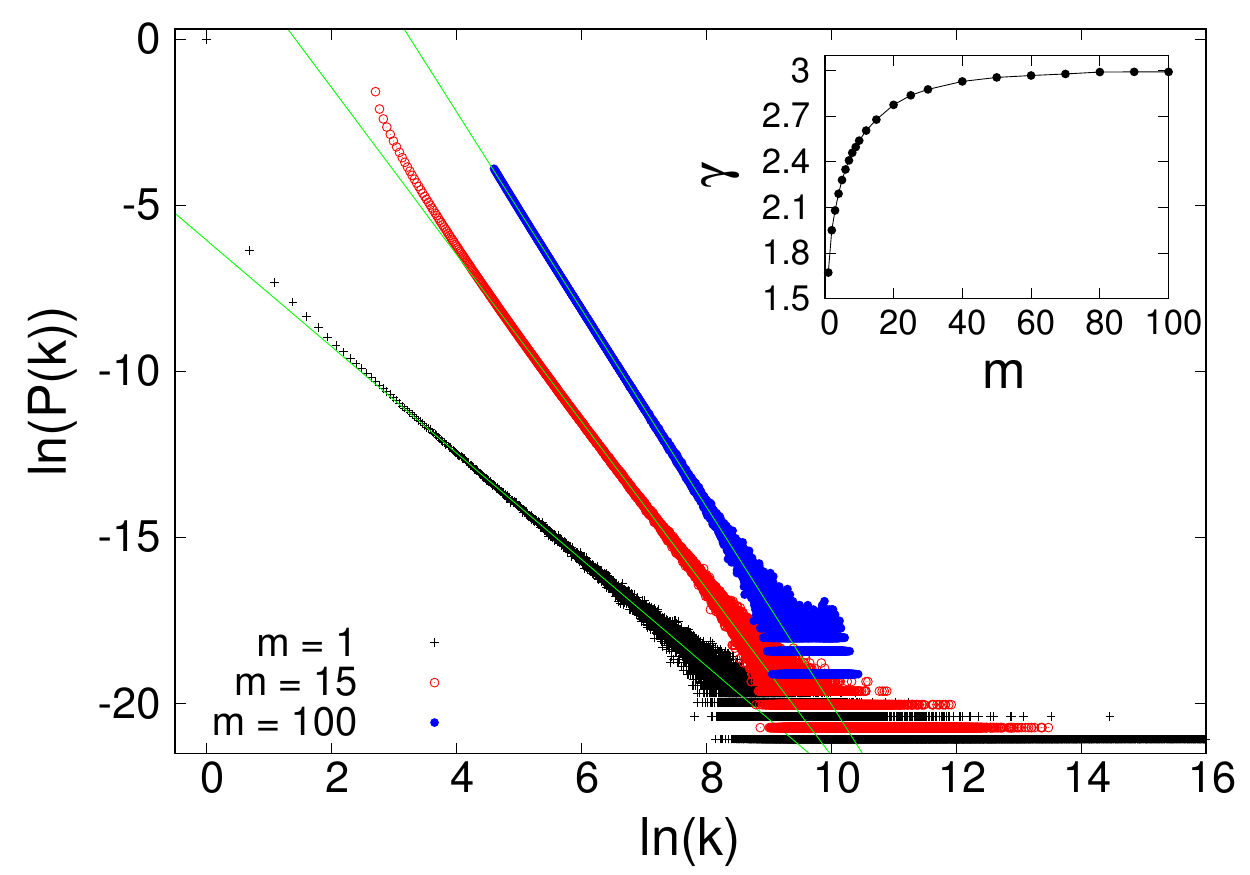}
\label{fig:2b}
}

\caption{
(a) Plots of $\ln(k_i)$ versus $\ln(t)$ for $5$ nodes added at five different times for
$m = 2$. In the inset, we show variation of the slope $\beta$ with $m$. (b) Plots of $\ln (P(k))$ vs $\ln(k)$ 
for $m = 1$, $m=15$ and $m=100$ revealing that the
MDA rule gives rise to power-law degree distribution. In the inset we show how the
exponent $\gamma$ approaches to its maximum value $3$ as $m\rightarrow \infty$.
} 

\label{fig:2ab}
\end{figure}

Solving Eq. (\ref{eq:3}) subject to the initial condition that the $i$th node is born 
at time $t=t_i$ with $k_i(t_i)=m$ gives,
\begin{equation} \label{eq:4}
 k_i(t) = m \Big ({\frac{t}{t_i}}\Big )^{\beta(m)}.
\end{equation}
The solution is exactly the same as that of the BA model except the fact the exponent $\beta$ is 
not a fixed value $\beta=1/2$ rather it depends on $m$ as we find $1/2\leq \beta<1$. To verify 
Eq. (\ref{eq:4}) for MDA model we plot $\log(k_i(t))$ versus $\log(t)$ in Fig. (\ref{fig:2a}) 
and find a set of straight lines with same slope regardless of the node we pick. It implies that they 
all gain connectivity following the same growth-law. The same MDA model has been studied by Yang {\it et al.} in 2013  \cite{ref.Yang}. 
They too gave a form for $\Pi(i)$ and resorted to
mean-field approximation. However, our expression for $\Pi(i)$ is totally different from
theirs and hence the value of $\beta$ we obtain. By invoking the
idea of cumulative probability and then following
the same procedure as has been done by Barab\'{a}si and Albert in their seminal paper,
we can immediately write
the solution for the degree distribution
\begin{equation} \label{eq:20}
 P(k) \sim k^{-\gamma(m)}, \: \text{where} \: \: \gamma(m) = \frac{1}{\beta(m)}+1. 
\end{equation}
One of the most significant difference of this result from that of the BA model is that the exponent
$\gamma$ depends on $m$ which we show in Fig. (\ref{fig:2b}). Note that  there is one point stands alone in 
the plot of degree distribution $P(k)$ for $m=1$. We find that almost $99$ percent of the total nodes
in any given realization have degree one. Data from single realization reveals that there is always one node which almost
always gains links. The MDA model for $m=1$ thus describes the {\it winner gets all} mechanism. 
However, as $m$ increases
{\it winner gets all} mechanism becomes weaker and for $m>12$ the stand alone point disappears completely and 
joins the trend of the mainstream data points.

\begin{figure}

\centering

\subfloat[]
{
\includegraphics[height=2.4 cm, width=4.0 cm, clip=true]
{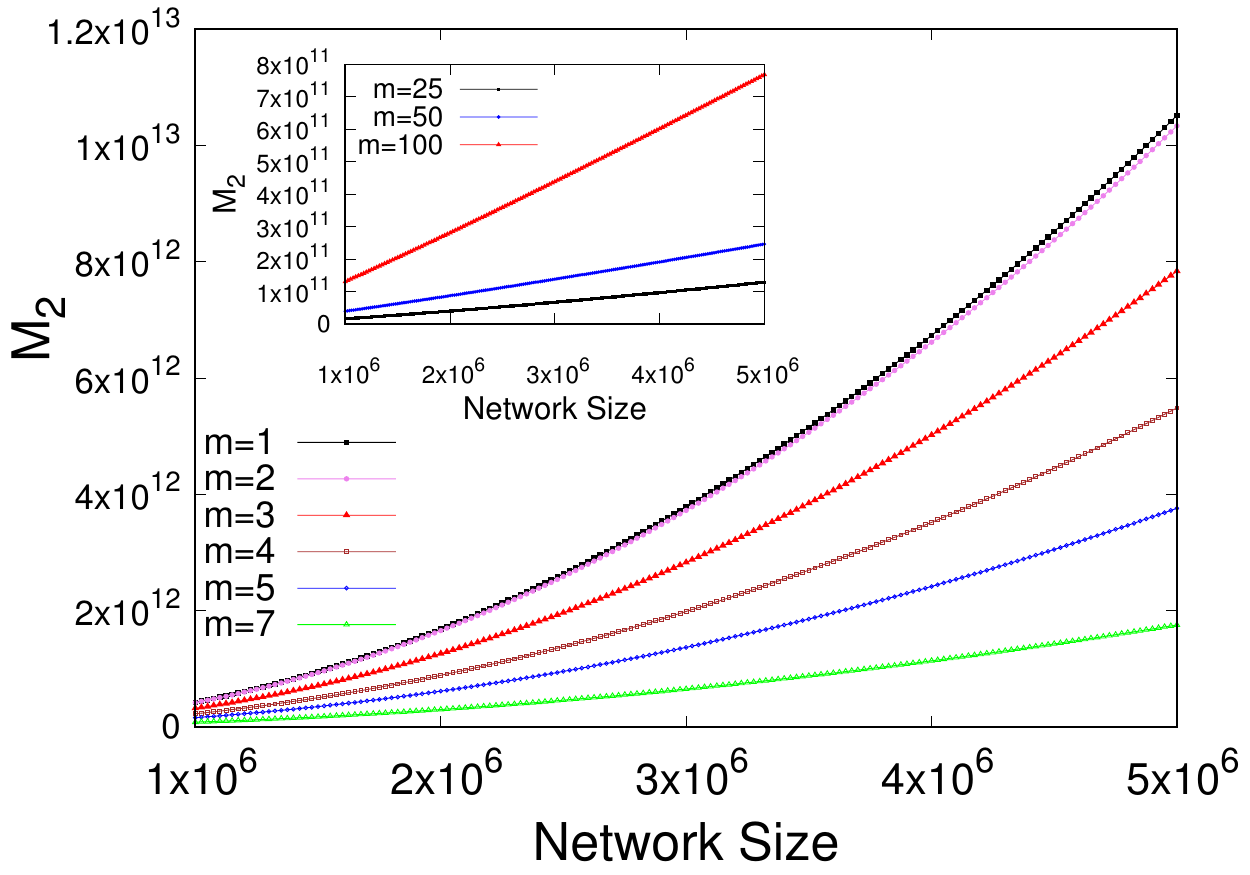}
\label{fig:3a}
}
\subfloat[]
{
\includegraphics[height=2.4 cm, width=4.0 cm, clip=true]
{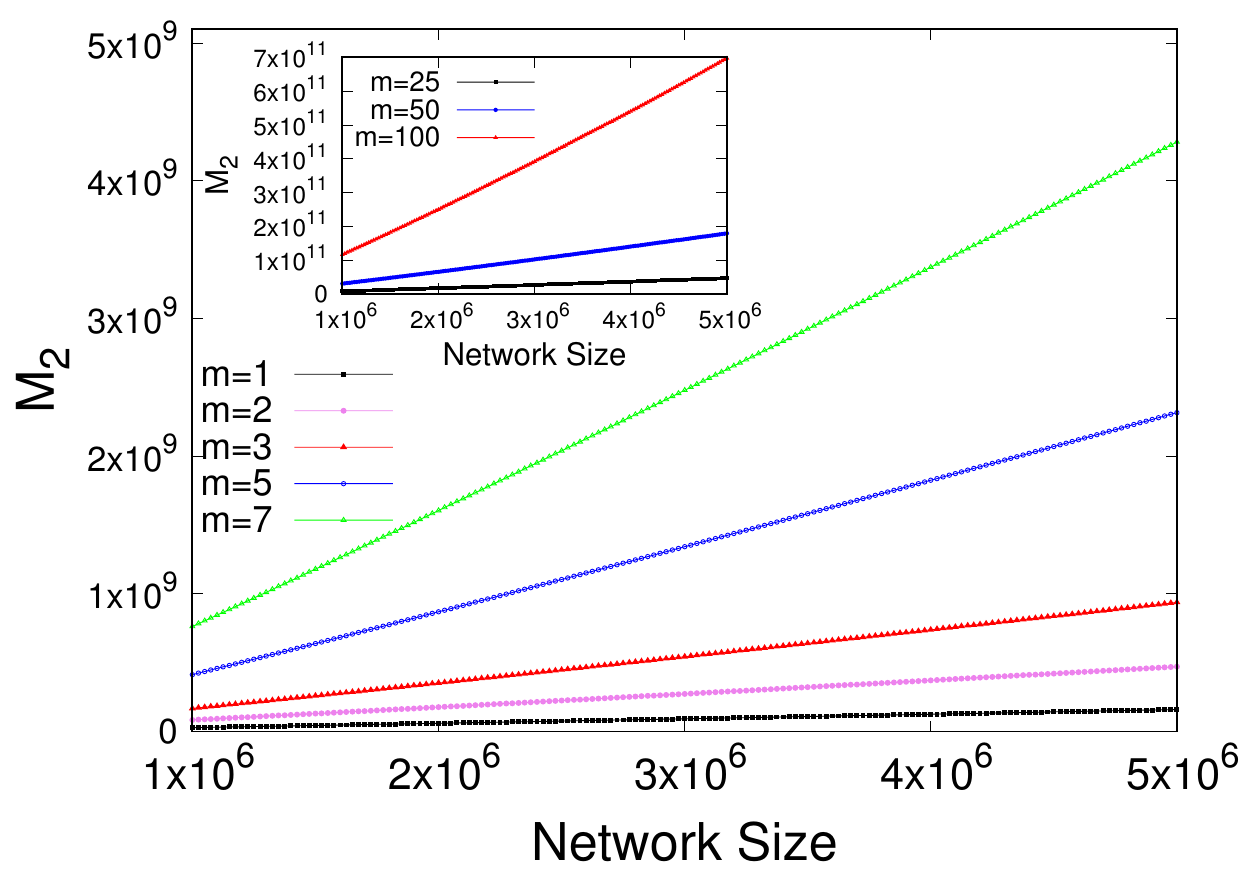}
\label{fig:3b}
}
\caption{ Plots of the second moment of the degree distribution for (a) MDA model and (b) BA model are shown for different $m$ to see the contrast.
} 
\label{fig:3ab}
\end{figure}

One of the useful measures of networks is the 
moment of the degree distribution \cite{ref.network_science, ref.Pastor}. The $n$th moment of the degree distribution is defined as
\begin{equation}
M_n=\int_{k_{{\rm min}}}^{k_{{\rm max}}} k^2P(k) dk.   
\end{equation}
For networks with power-law degree distribution with $\gamma<3$, the first moment is finite but the second
moment may diverge. The divergence of $M_2$ for large network size indicates that the fluctuations around
the average degree can be arbitrary large. It implies that when we pick a node at random, 
we cannot predict the degree of the selected node. Its value can be small or arbitrarily large. Thus networks with $\gamma<3$ do
not have a meaningful internal scale and hence it is called scale-free. In this sense, our
MDA network truly is a scale-free network. Strictly speaking $\langle k^2\rangle$
 diverges only in the $N \rightarrow \infty$ limit. Yet, the divergence
is relevant for finite networks as well. In Fig. (\ref{fig:3ab}) we show plots of the second moment 
$M_2$ for both MDA and BA model. It is clear that in the case of the MDA model, the value of $M_2$ increases
with the network size $N$. 
However, we find that for small $m$, the growth of $M_2$ is non-linear and as $m$ increases
the extent of non-linearity decreases so that for $m>50$ it is almost linear. The higher the
non-linearity means that there are fewer but richer hubs and the disparity between the rich and the 
poor is higher.
On the other hand, for BA model, $M_2$ is always linear with $N$ though their slopes increases with increasing $m$.
This is because as $m$ increases the average degree increases. We observe that for large $m$, such as for $m$ beyond $50$, the $M_2$ versus $N$ curves of the two models starts to get so close that at $m=100$ they are
almost the same. Note that 
\begin{equation}
    {{kP(k)}\over{\sum_k kP(k)}}={{kP(k)}\over{\langle k\rangle}},
\end{equation}
is a quantity which describes the probability that a node picked at random has a neighbour
whose degree is exactly $k$. Then the quantity
\begin{equation}
    \sum k{{kP(k)}\over{\sum_k kP(k)}}={{\langle k^2 \rangle}\over{\langle k\rangle}},
\end{equation}
gives a measure of the average degree of a neighbour. If this quantity is greater than $\langle k \rangle$ then
it means that the neighbours of a node have more degrees than the node itself. Thus the two models
differ significantly more for small $m$ and eventually start to be similar in the large $m$ value.


\section{Dynamic Scaling}

Interestingly, although the nodes gain links preferentially yet the degree of all the nodes 
in a given realization grows following the same growth-law $k\sim t^\beta$ with the same exponent
$\beta$ for a given value of $m$ (see  Fig. (\ref{fig:2a})). The rest of our analysis 
is based on the solution for $k_i(t)$ given
by Eq. (\ref{eq:4}). We find it highly instructive to combine the two variables, the degree $k_i$ 
of the $i$th node and its birth time $t_i$ 
into a single one by using of  Eq. (\ref{eq:4}). It suggests that a node say $i$ at time $t$ can be
better characterized by the generalized degree
\begin{equation} \label{eq:qiti}
 q_{i}(t) = {{k_{i}}\over{m}} t_{i}^{\beta},
\end{equation}
since a node which is born now cannot be compared on equal footing with the one  born
earlier. The generalized degree combines the degree and birth time in such a way
that a node which has high degree is highly likely to have a low birth time. On the other hand,
nodes with low degree value definitely have high birth time and hence has relatively much
less time to acquire high degree. According to Eq. (\ref{eq:4}) we find that the generalized degree grow with 
time as
\begin{equation}\label{eq:qtbeta}
 q_{i}(t) \sim t^{\beta}. 
\end{equation}
It implies that instead of characterizing nodes by their degree $k$ we can characterize
them by their generalized degree $q$ and hence the corresponding network by
the generalized degree distribution $F(q,t)$.

\begin{figure}

\centering

\subfloat[]
{
\includegraphics[height=2.4 cm, width=4.0 cm, clip=true]
{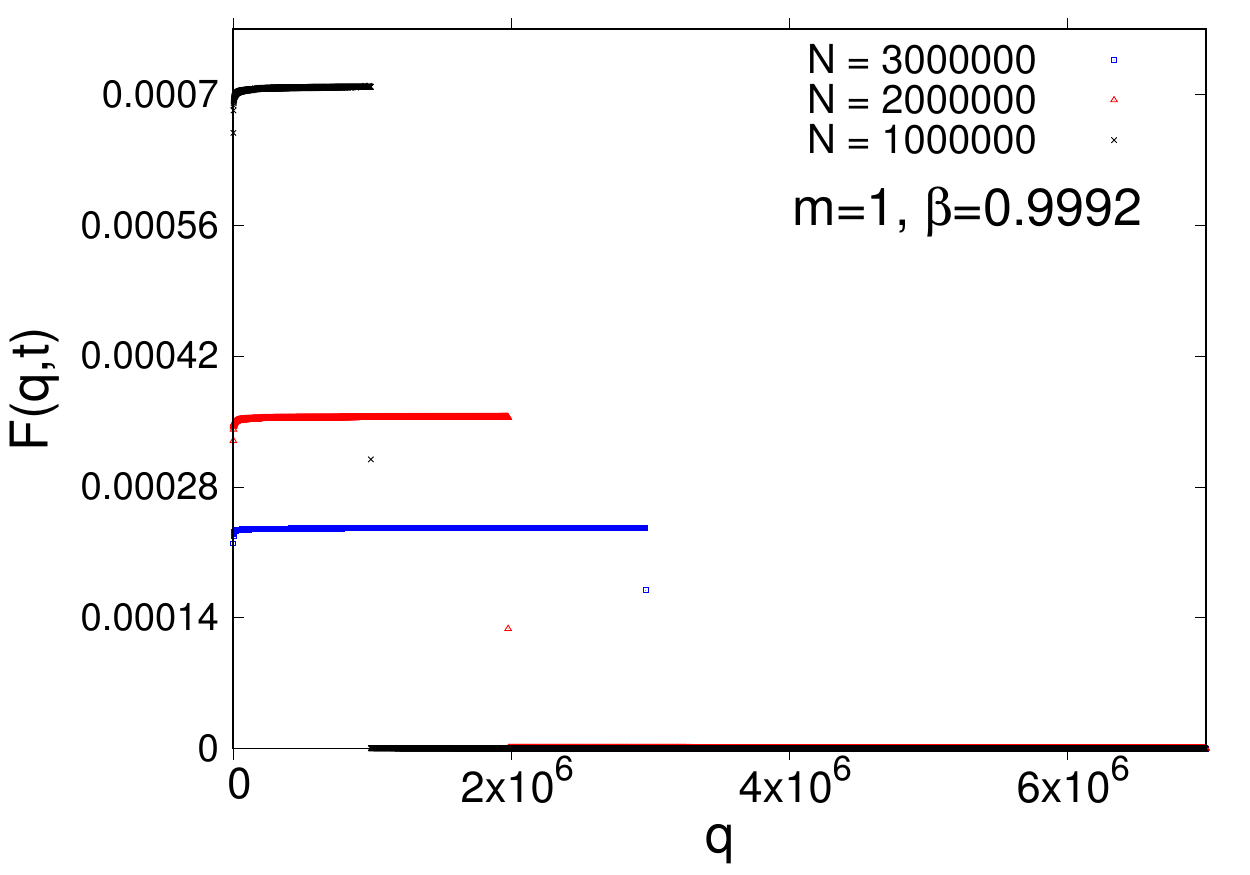}
\label{fig:4a}
}
\subfloat[]
{
\includegraphics[height=2.4 cm, width=4.0 cm, clip=true]
{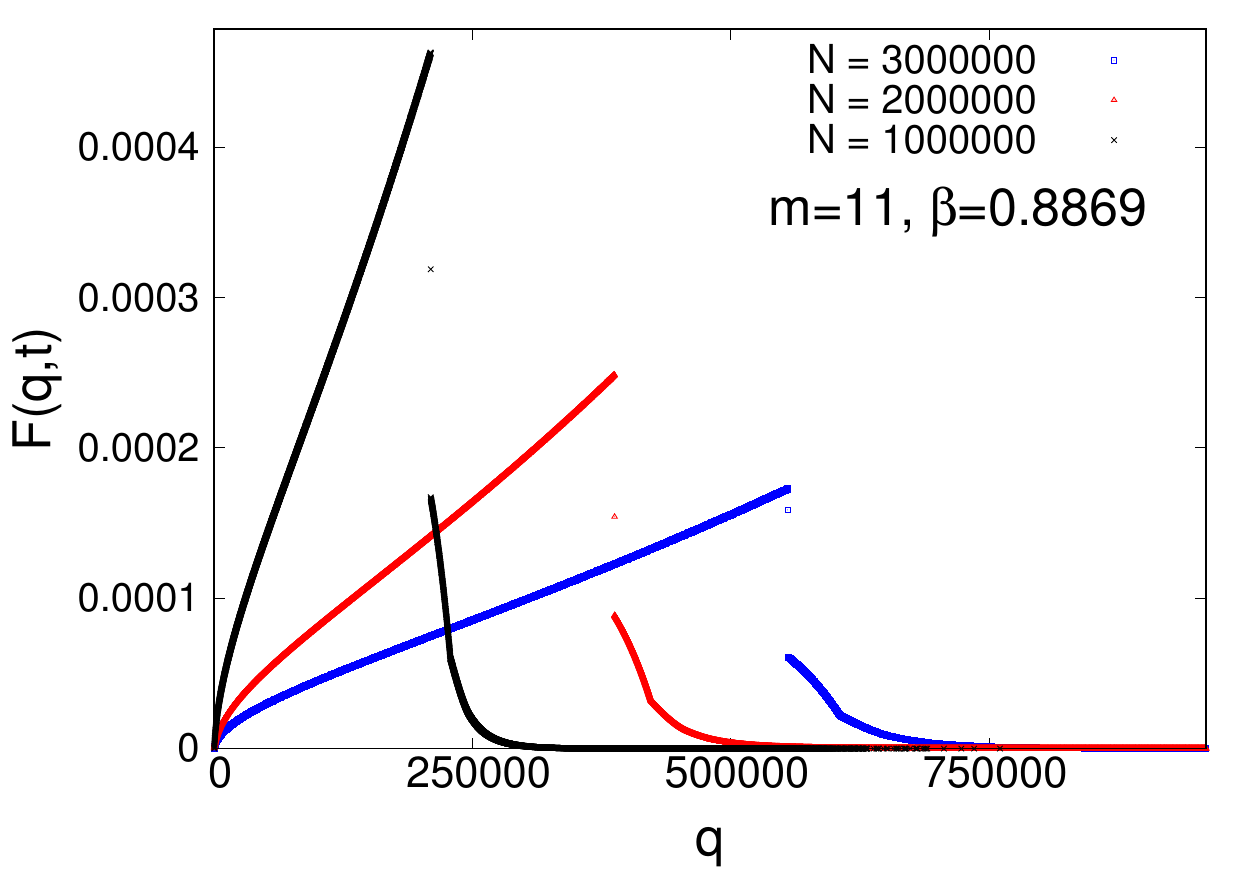}
\label{fig:4b}
}

\subfloat[]
{
\includegraphics[height=2.4 cm, width=4.0 cm, clip=true]
{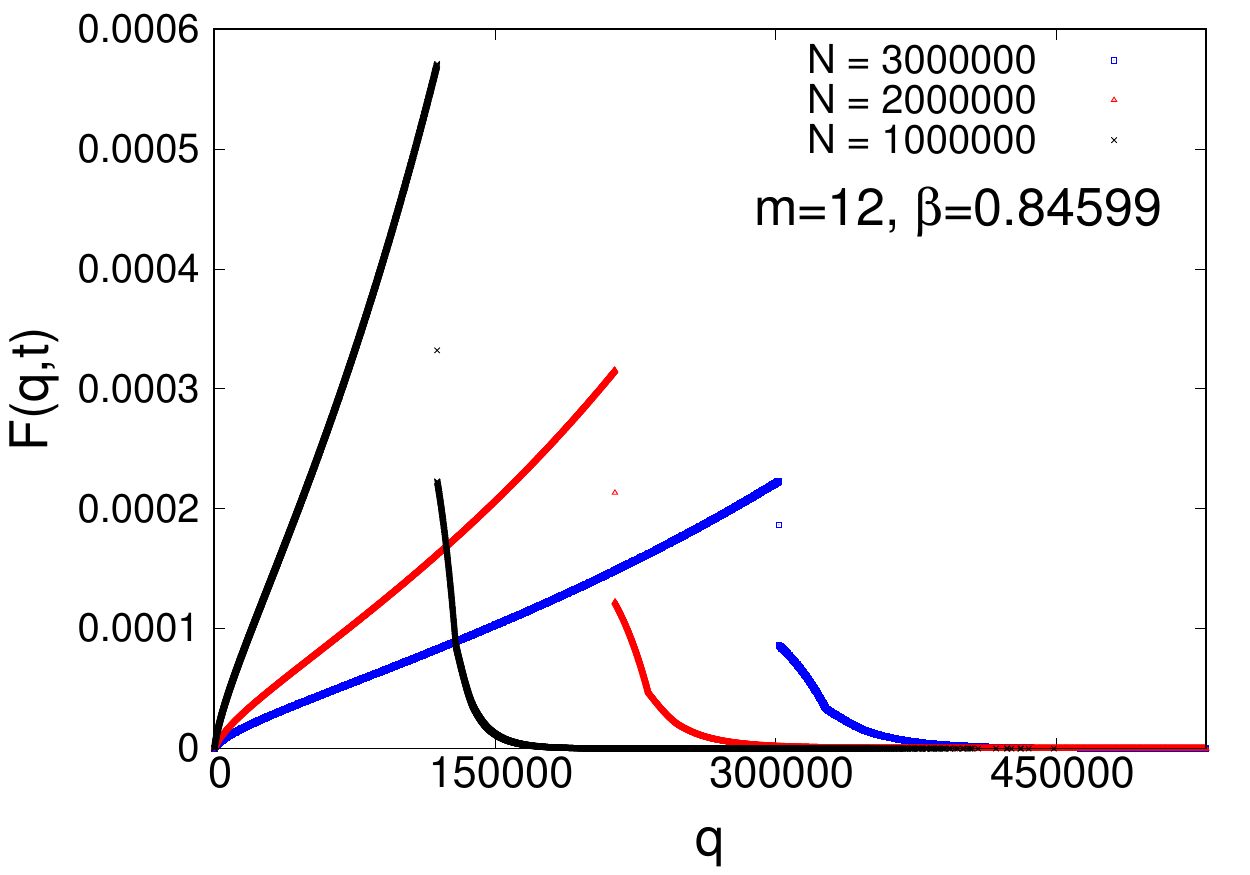}
\label{fig:4c}
}
\subfloat[]
{
\includegraphics[height=2.4 cm, width=4.0 cm, clip=true]
{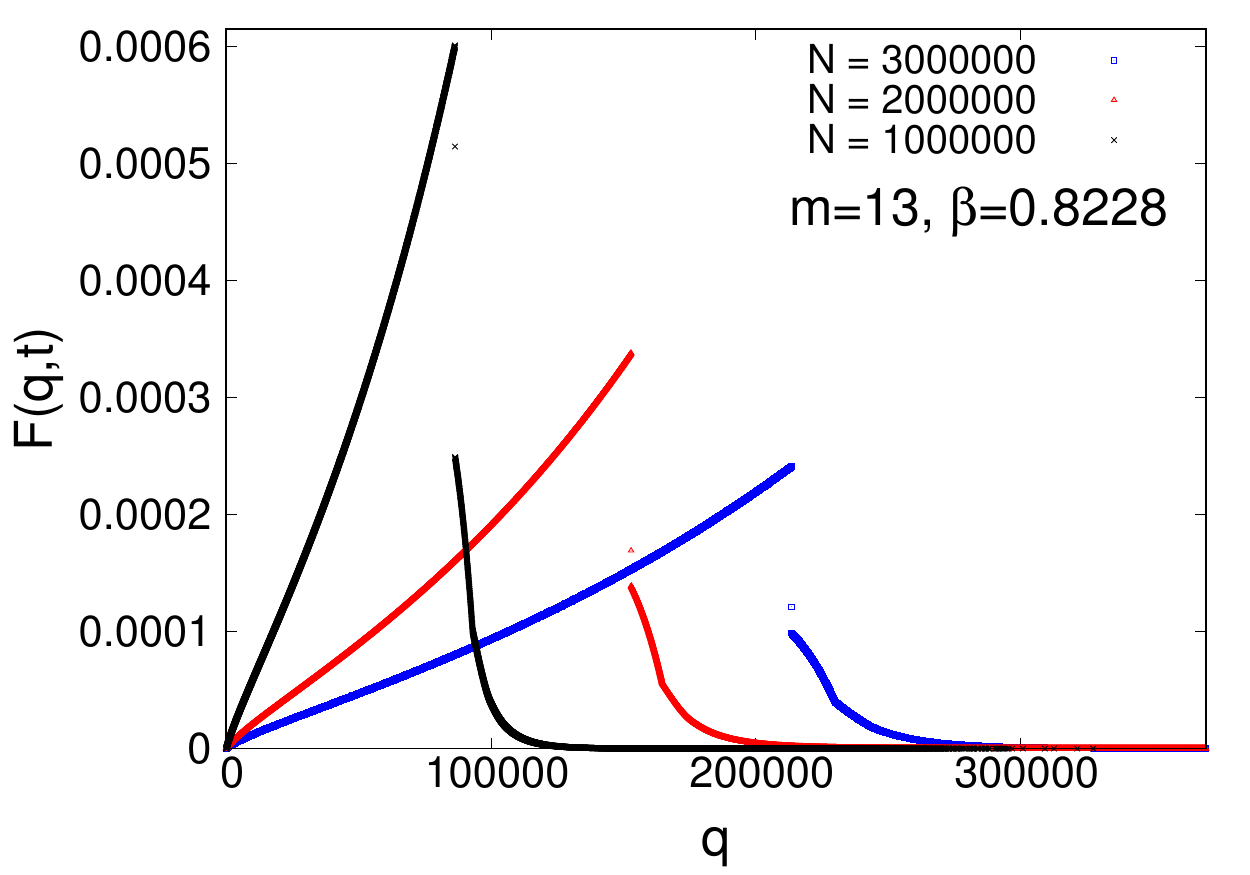}
\label{fig:4d}
}

\subfloat[]
{
\includegraphics[height=2.4 cm, width=4.0 cm, clip=true]
{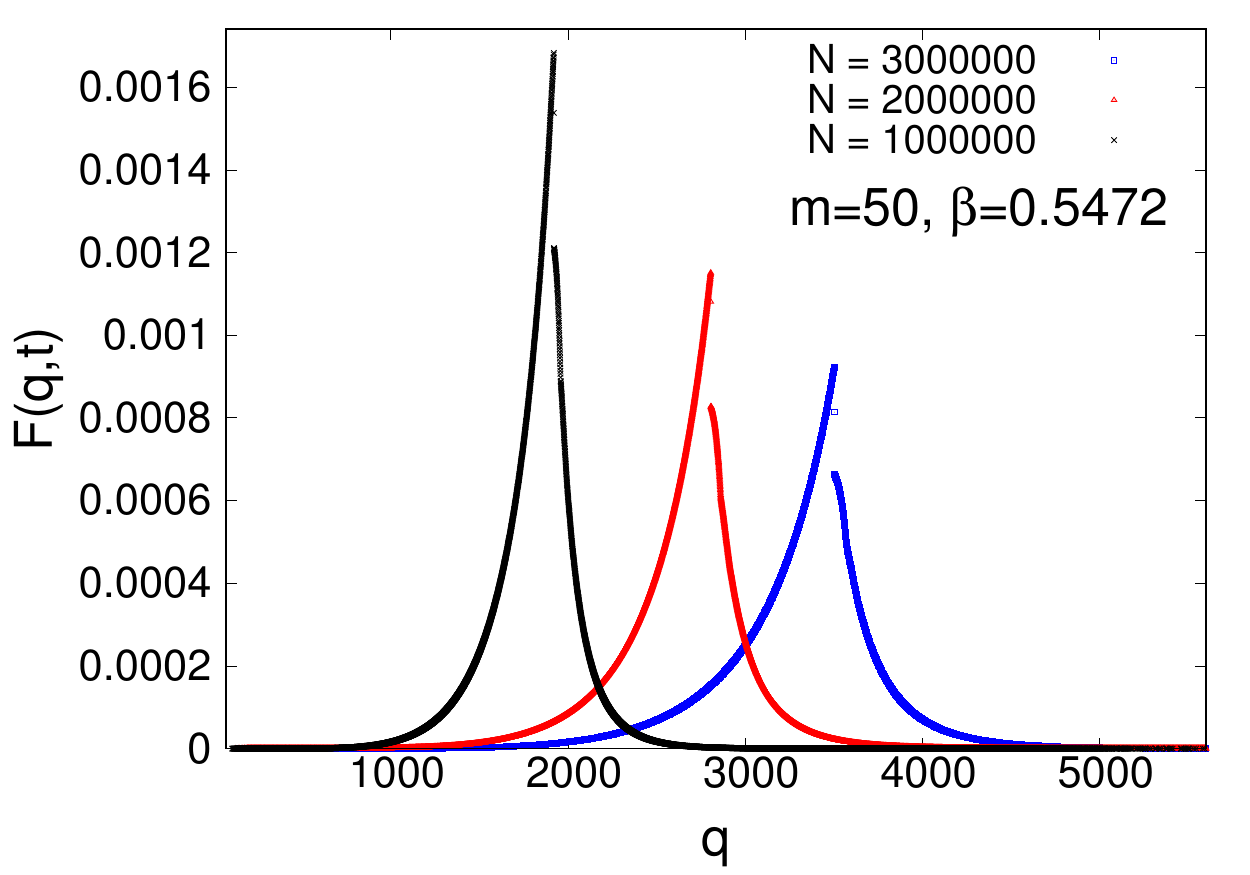}
\label{fig:4e}
}
\subfloat[]
{
\includegraphics[height=2.4 cm, width=4.0 cm, clip=true]
{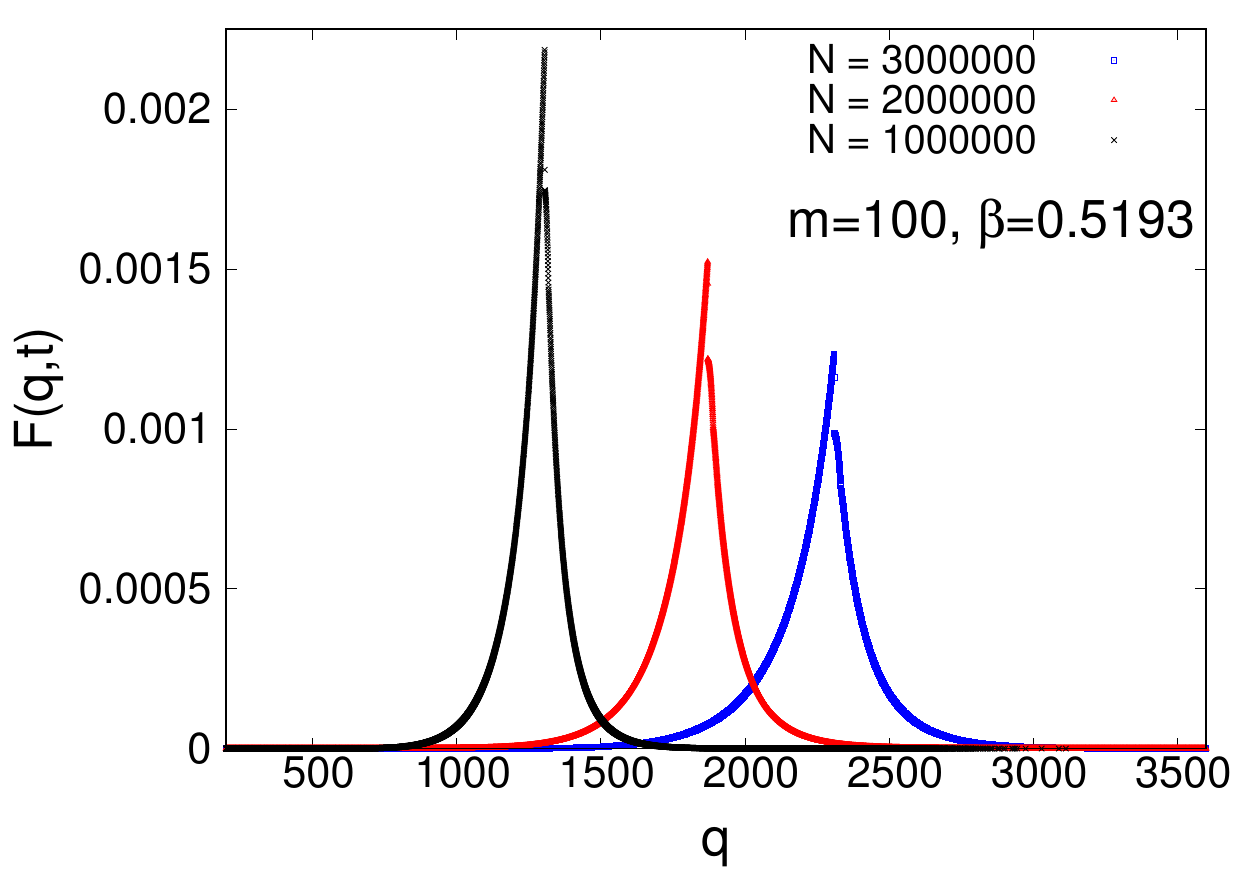}
\label{fig:4f}
}

\caption{
Plots of the generalized degree distribution $F(q,t)$ vs $q$ of the MDA model for (a) $m=1$,(b)$m=11$, 
(c) $m=12$, (d) $m=13$, (e) $m=50$ and (f) $m=100$. It clearly demonstrate that the distribution of 
$F(q,t)$ is highly sensitive
to small $m$. However, we hardly see any difference between the generalized degree distribution 
$F(q,t)$ for $m=50$ and $m=100$.
} 

\label{fig:4abcdef}
\end{figure}

The generalized degree distribution $F(q,t)$ is defined as the probability that
a node picked at random has generalized degree $q$ at time $t$. We performed
extensive Monte Carlo simulation and collected ensemble averaged data over many 
independent realizations for a fixed time to obtain $F(q,t)$ as a function of $q$.
Some of the plots of $F(q,t)$ as a function of $q$ for different $m$ and $N$
 are shown in Fig. (\ref{fig:4abcdef}). 
 It is interesting to see how the layout of the plots change with $m$. 
The summary of the trend seen in the plots are as follows. First, $m=1$ case is distinctly different from the rest. It consists of almost two horizontal lines
separated by a huge gap between rich and poor in $q$ value.  The same we have seen in Fig. (\ref{fig:2b}) for the degree distribution
where there is one point that stands alone from the rest. However, in all plots of $F(q,t)$ 
versus $q$ for $m>1$, we find that $F(q,t)$ rises 
in a concave fashion for low $1<m<12$  and in a convex manner for $m>12$. Second, 
there is a sudden drop from the peak to some non-zero 
value at $q = q_c$ followed by a decrease in a concave manner which then
decay along a long tail. Third, the gap size of the drop decreases with increasing
$m$. Fourth, as $N$ is increased, the value of 
$q_c$, exactly from where drop occurs, increases at the expense of decreasing height. 
Fifth, we find our mean-field approximation works well at $m>12$
as the attachment probability becomes
proportional to $k$ like the PA rule of the BA model. In Ref. \cite{ref.hassan_liana},
we have seen that the inverse
 harmonic mean (IHM) of the degrees of the mediators of each node 
 fluctuate wildly for $m<12$ and hence the mean of the
IHM value of all the nodes bears no meaning. However, as $m$ increases mean of IHM starts to have 
meaning as the fluctuation decreases with increasing  $m$ above $m\approx 15$.

\begin{figure}

\centering

\subfloat[]
{
\includegraphics[height=2.4 cm, width=4.0 cm, clip=true]
{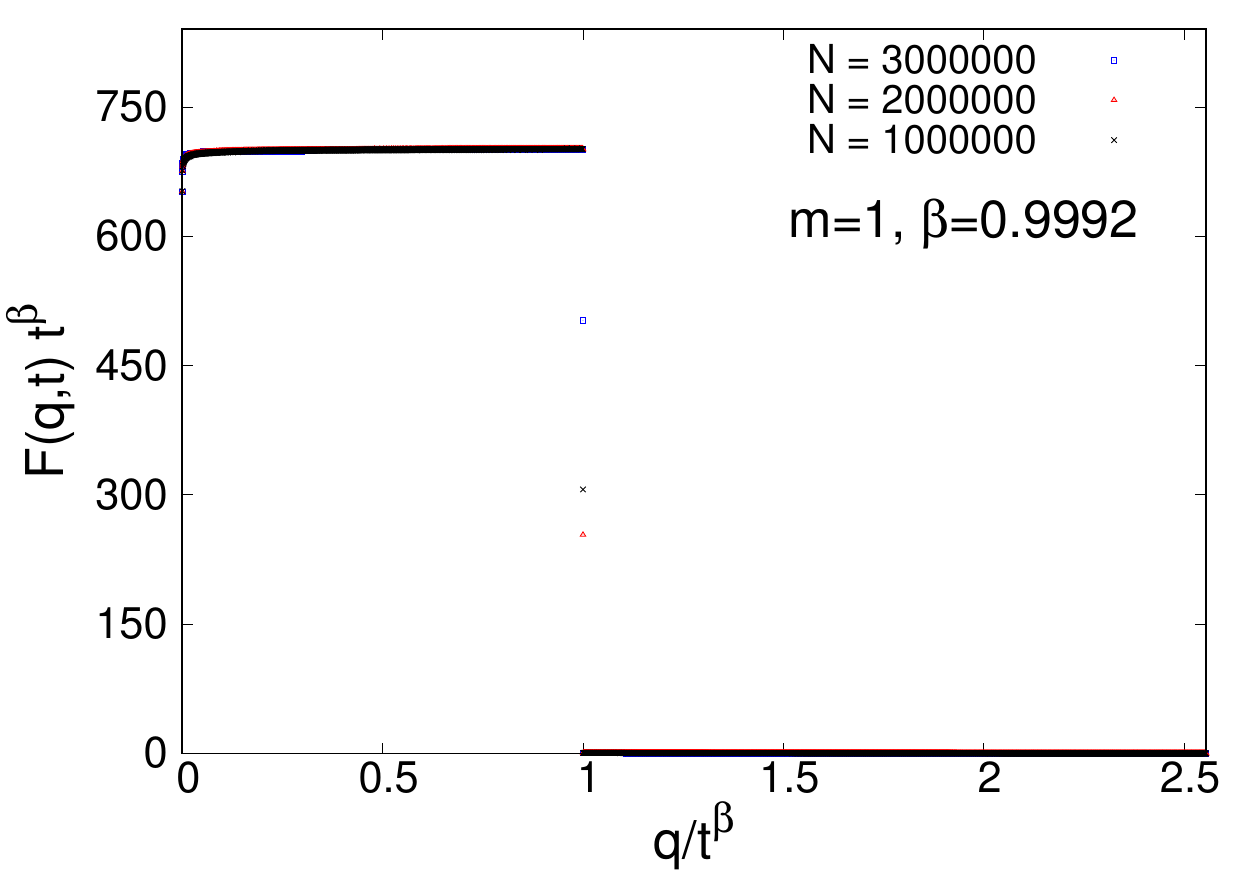}
\label{fig:5a}
}
\subfloat[]
{
\includegraphics[height=2.4 cm, width=4.0 cm, clip=true]
{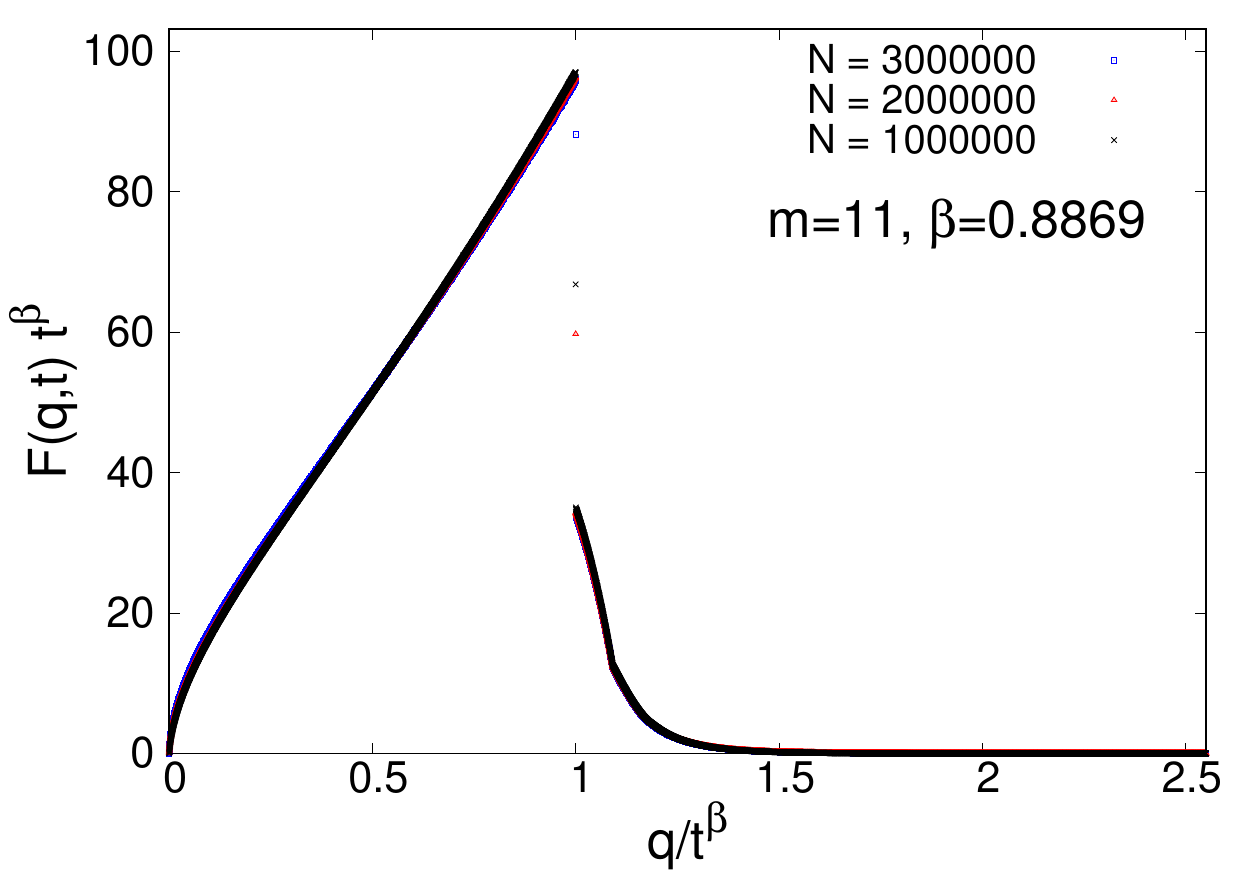}
\label{fig:5b}
}

\subfloat[]
{
\includegraphics[height=2.4 cm, width=4.0 cm, clip=true]
{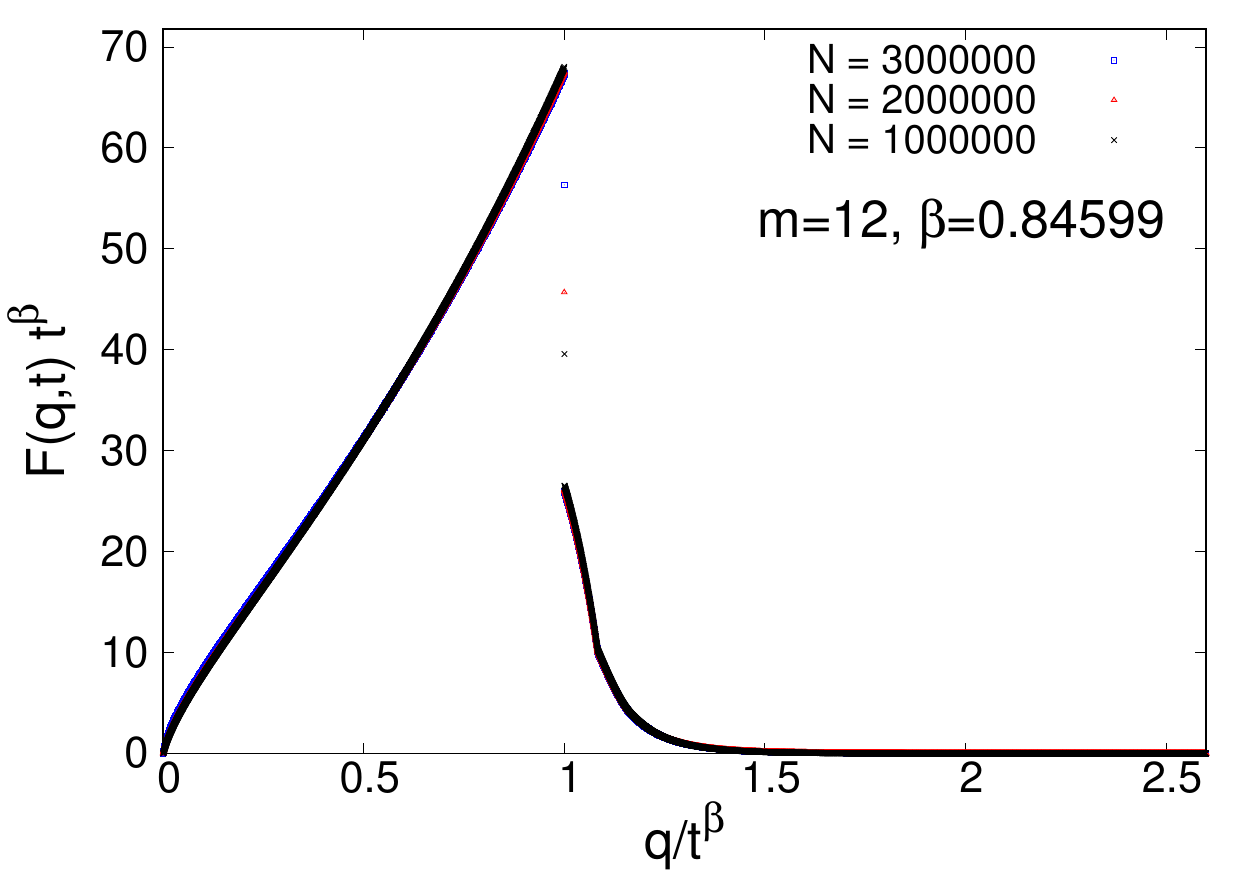}
\label{fig:5c}
}
\subfloat[]
{
\includegraphics[height=2.4 cm, width=4.0 cm, clip=true]
{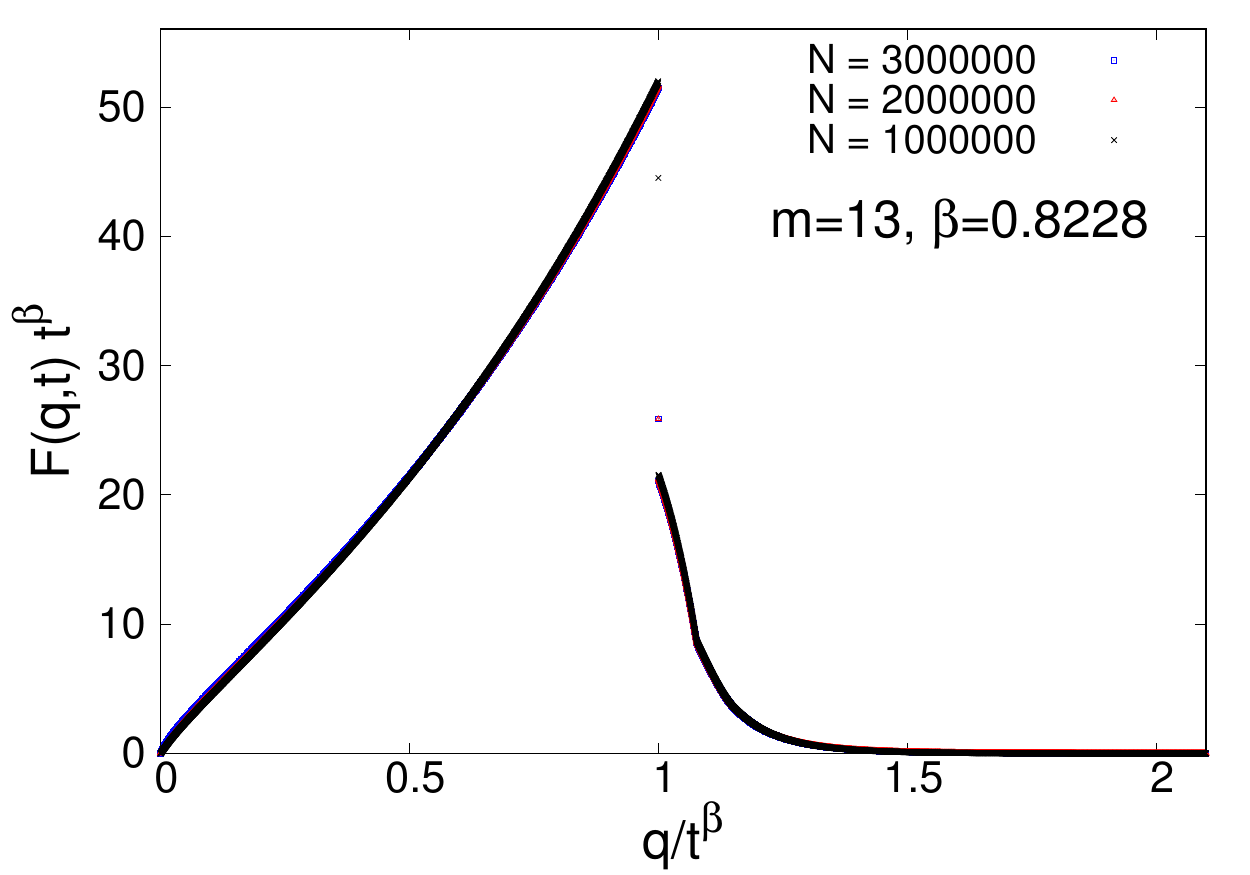}
\label{fig:5d}
}

\subfloat[]
{
\includegraphics[height=2.4 cm, width=4.0 cm, clip=true]
{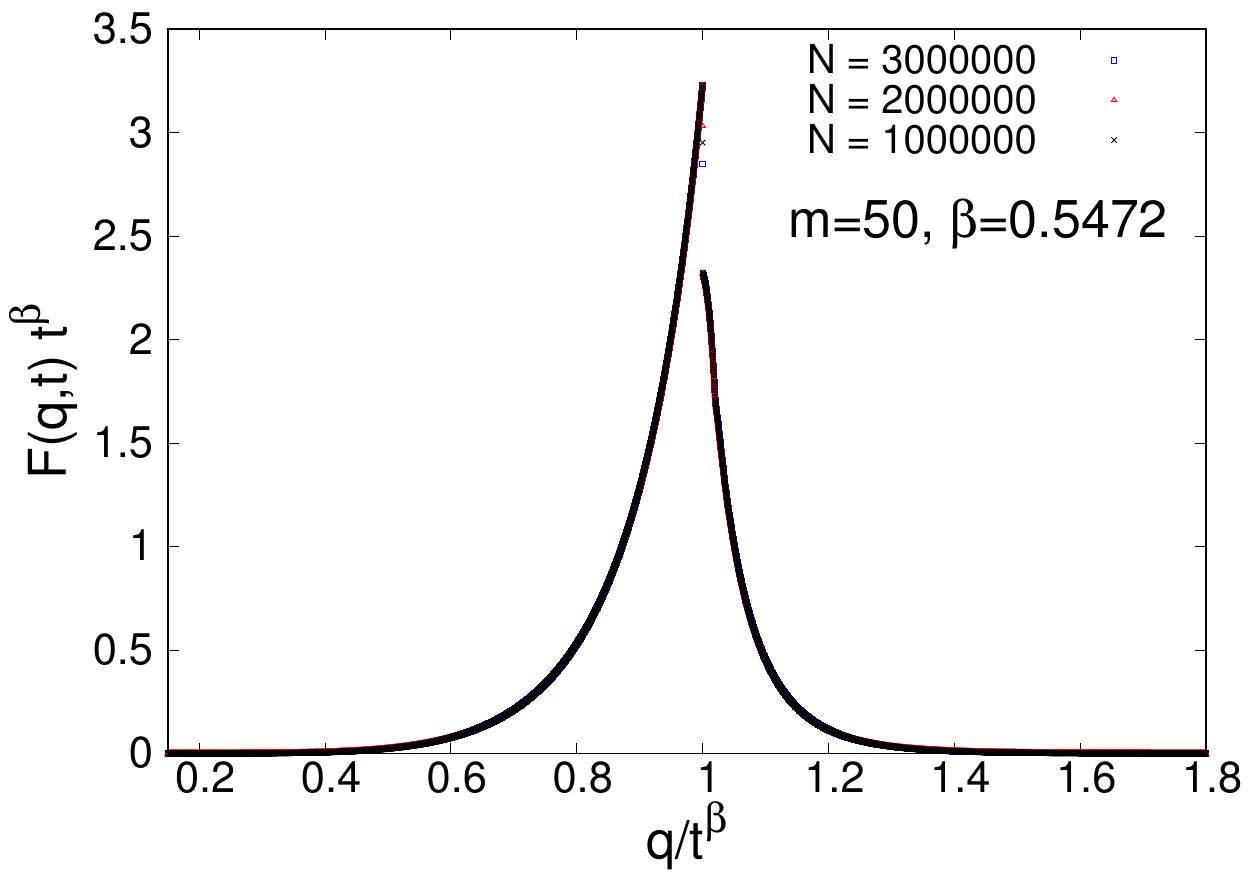}
\label{fig:5e}
}
\subfloat[]
{
\includegraphics[height=2.4 cm, width=4.0 cm, clip=true]
{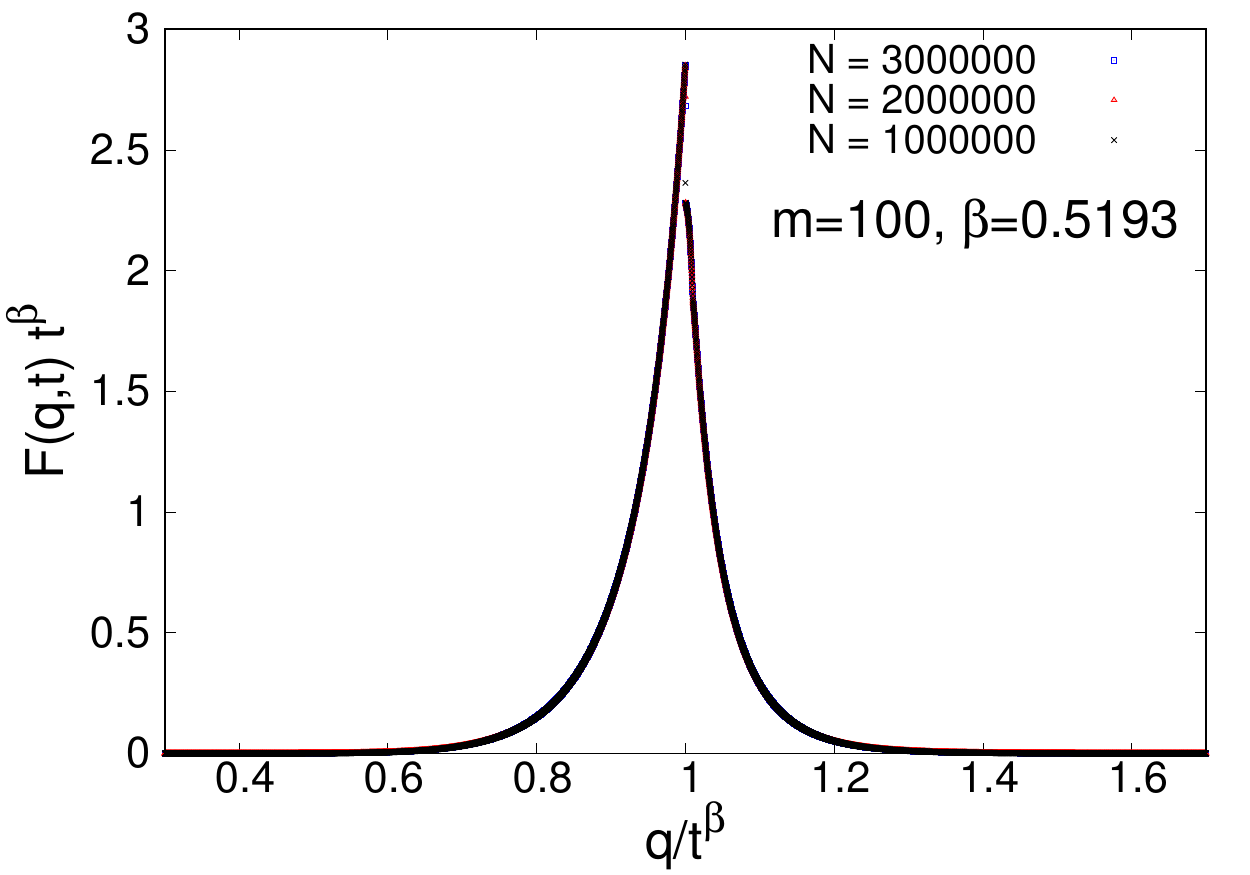}
\label{fig:5f}
}

\caption{
We use the same data as in Fig. (\ref{fig:4abcdef}) for MDA model except the fact that $F(q,t)$ is now measured in
units of $t^{-\beta}$ and $q$ in units of $t^\beta$. These plots are shown in (a) to (f) for $m=1,11, 12,13,
50, 100$ and find that for each value of $m$ the distinct plots in Fig. (\ref{fig:4abcdef}) collapse superbly
proving that MDA networks grow following dynamic scaling.
} 

\label{fig:5abcdef}
\end{figure}

It is clear from Eq. (\ref{eq:qiti}) that there are two governing parameters $q$ and $t$, and one governed parameter $F(q,t)$ 
in the problem at hand. Inspired
by equation \eqref{eq:qtbeta}, we can define a dimensionless governing parameter
\begin{equation}
 \xi = \frac{q}{t^{\beta}}.
\end{equation}
Thus $F(q,t)$ too can be expressed in terms of $t$ alone because time has been chosen to 
be the independent
parameter in this situation. The dimension function of a physical quantity must be a power monomial
$[F(q,t)] = [t^{\theta}]$. We can thus define a dimensionless governed parameter
\begin{equation}
\label{eq:dimensionless_governed}
 \phi \sim \frac{F(q,t)}{t^{\theta}}.
\end{equation}
The quantity $\phi$ is a dimensionless governed quantity and hence its numerical value
can only depend on the dimensionless governing parameter $\xi$.
Using it in Eq. (\ref{eq:dimensionless_governed}) we can write
\begin{equation}
\label{eq:dynamic_scaling}
 F(q,t) \sim t^{\theta} \phi \left(\frac{q}{t^{\beta}}\right),
\end{equation}
where the value of $\theta$ can be determined by using the normalization condition
$\int_0^\infty F(q,t) \, \mathrm{d} q = 1$ which gives $\theta = - \beta$. 
An observable quantity $f(x,t)$ that depends on two variables of which one is time $t$ and takes
the form 
\begin{equation}
  f(x,t)\sim t^\alpha\phi(x/t^z),
\end{equation}
is said to exhibit dynamic scaling \cite{ref.family}.
It implies that we can obtain the distribution function $F(q,t)$ at different 
moments of time by a similarity transformation
\begin{equation}
 q \longrightarrow \lambda^{\beta} q,\ t \longrightarrow \lambda t,\ F \longrightarrow \lambda^{-\beta}F.
\end{equation}
It is thus expected that the function $F(q,t)$ to be temporally self-similar. So we looked for 
data-collapse using
numerical simulations which is the best way to verify whether the system really exhibits
dynamic scaling or not.

\begin{figure}

\centering

\subfloat[]
{
\includegraphics[height=2.4 cm, width=4.0 cm, clip=true]
{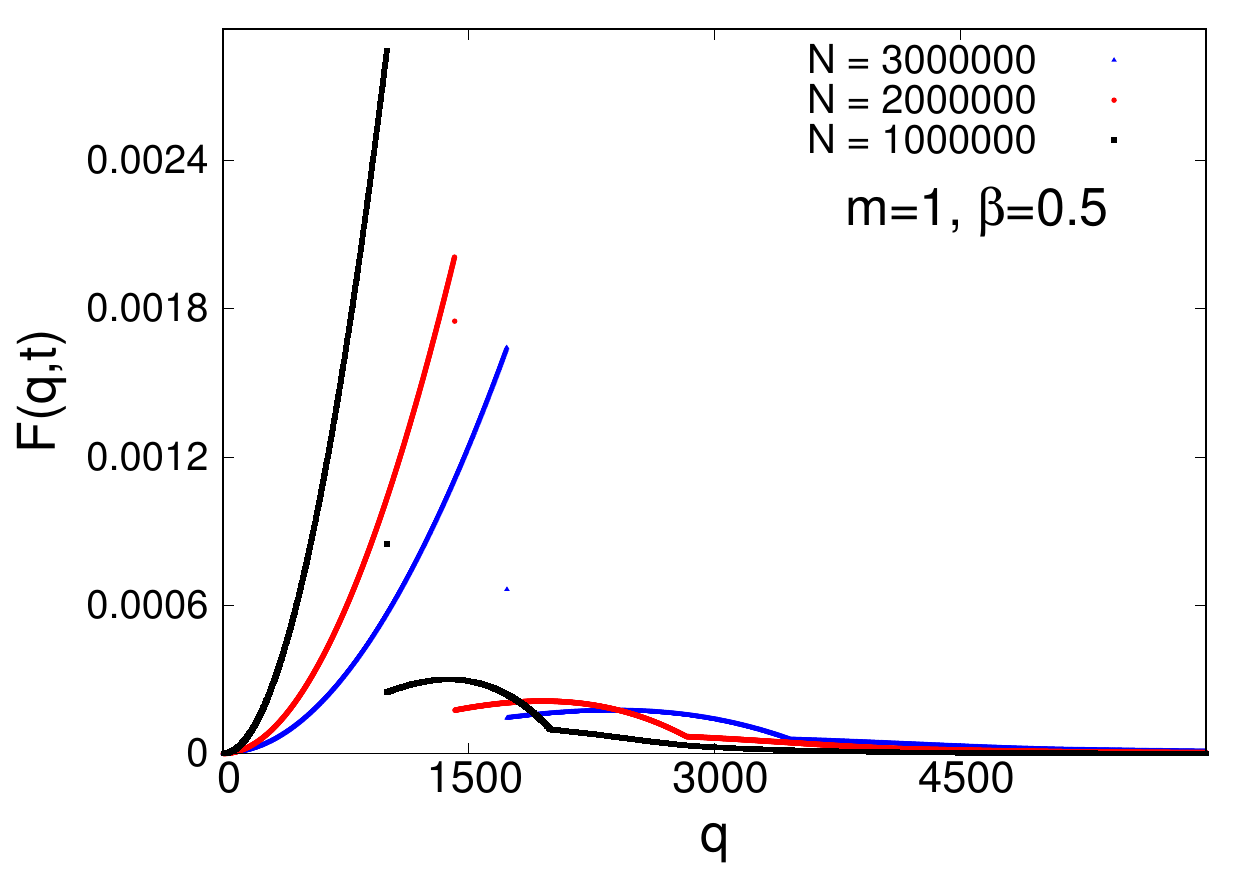}
\label{fig:6a}
}
\subfloat[]
{
\includegraphics[height=2.4 cm, width=4.0 cm, clip=true]
{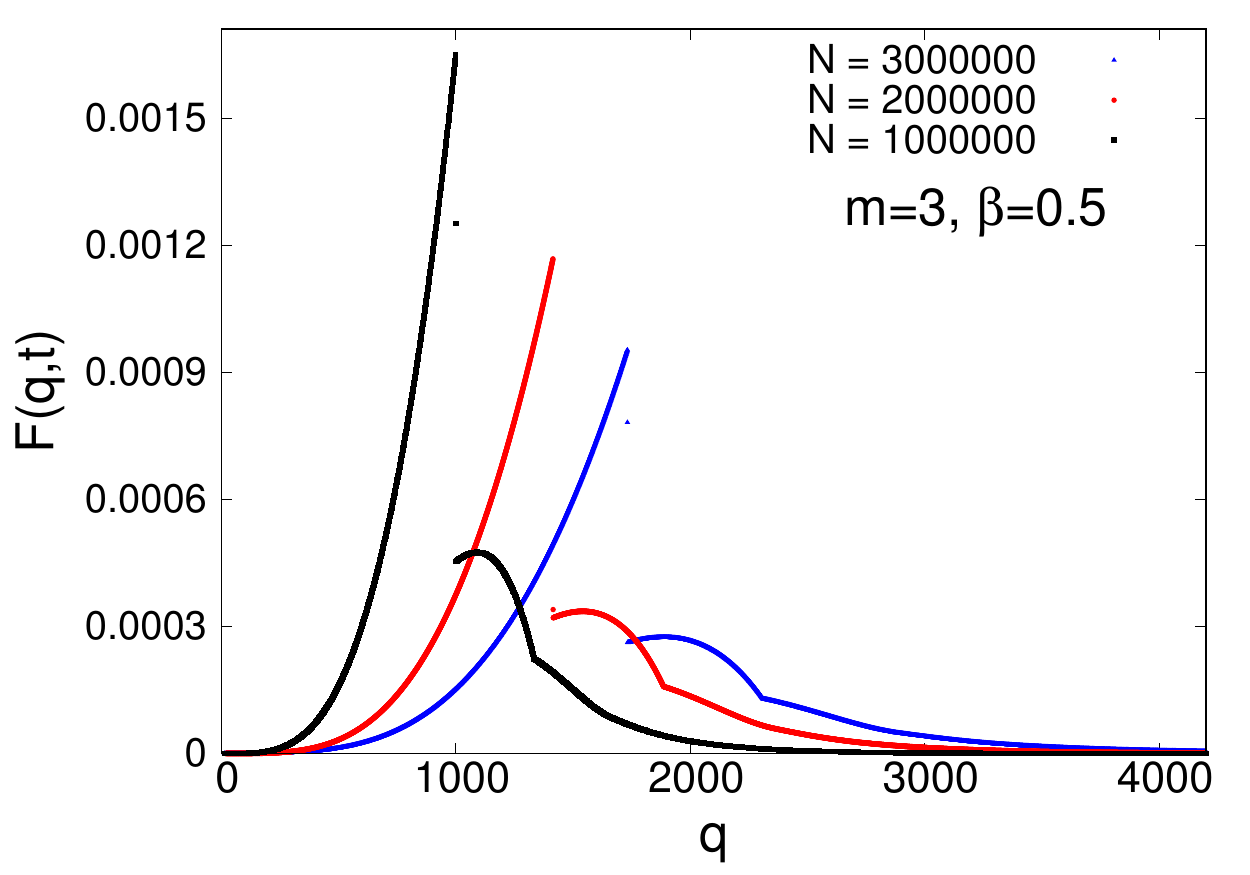}
\label{fig:6b}
}

\subfloat[]
{
\includegraphics[height=2.4 cm, width=4.0 cm, clip=true]
{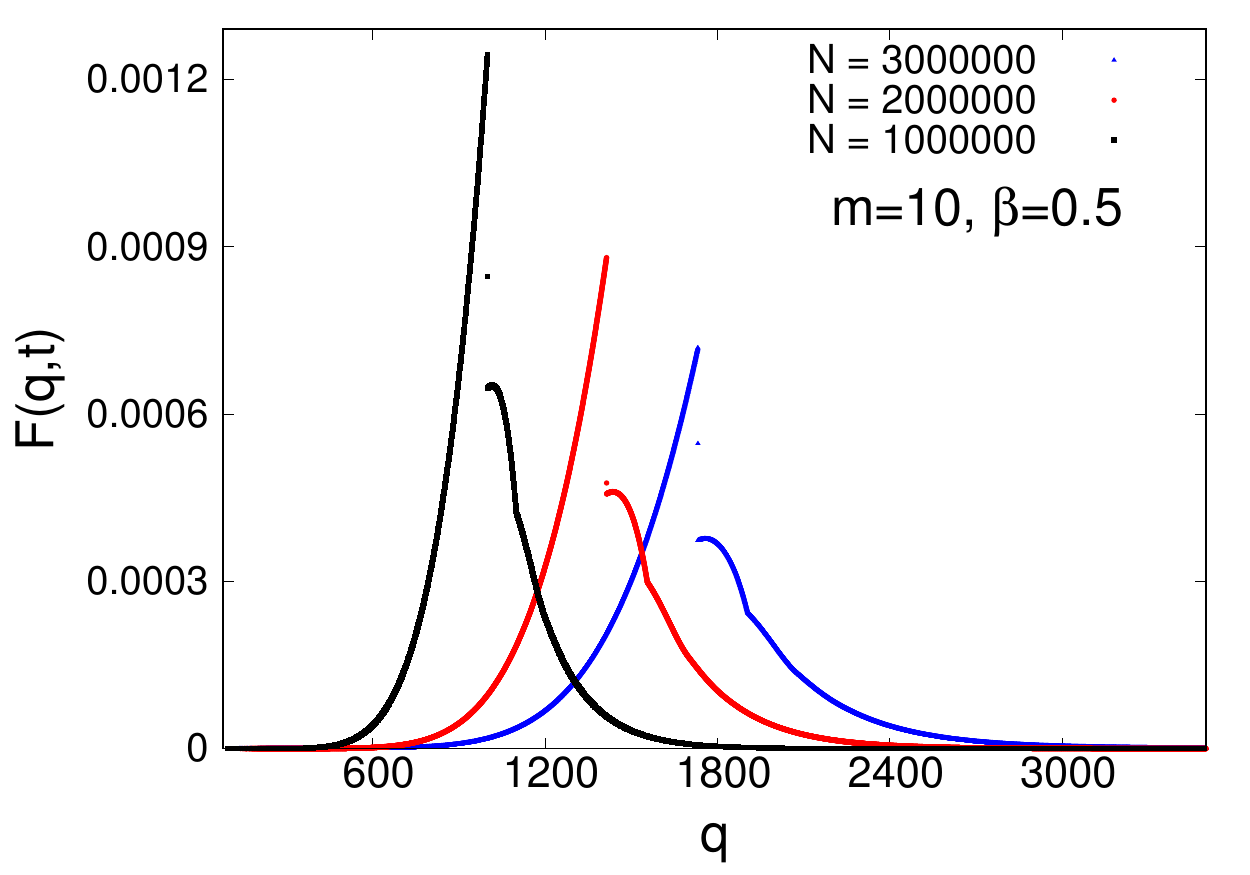}
\label{fig:6c}
}
\subfloat[]
{
\includegraphics[height=2.4 cm, width=4.0 cm, clip=true]
{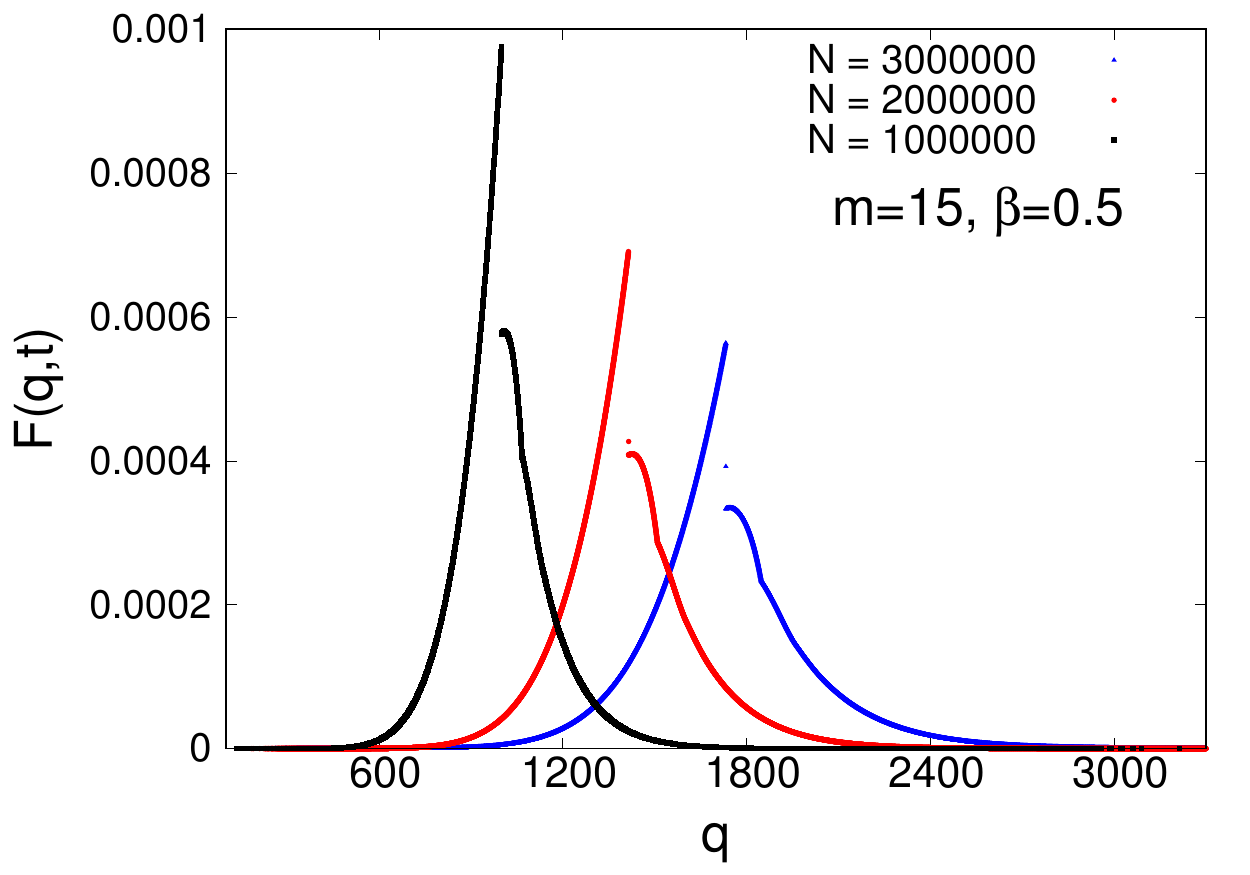}
\label{fig:6d}
}

\subfloat[]
{
\includegraphics[height=2.4 cm, width=4.0 cm, clip=true]
{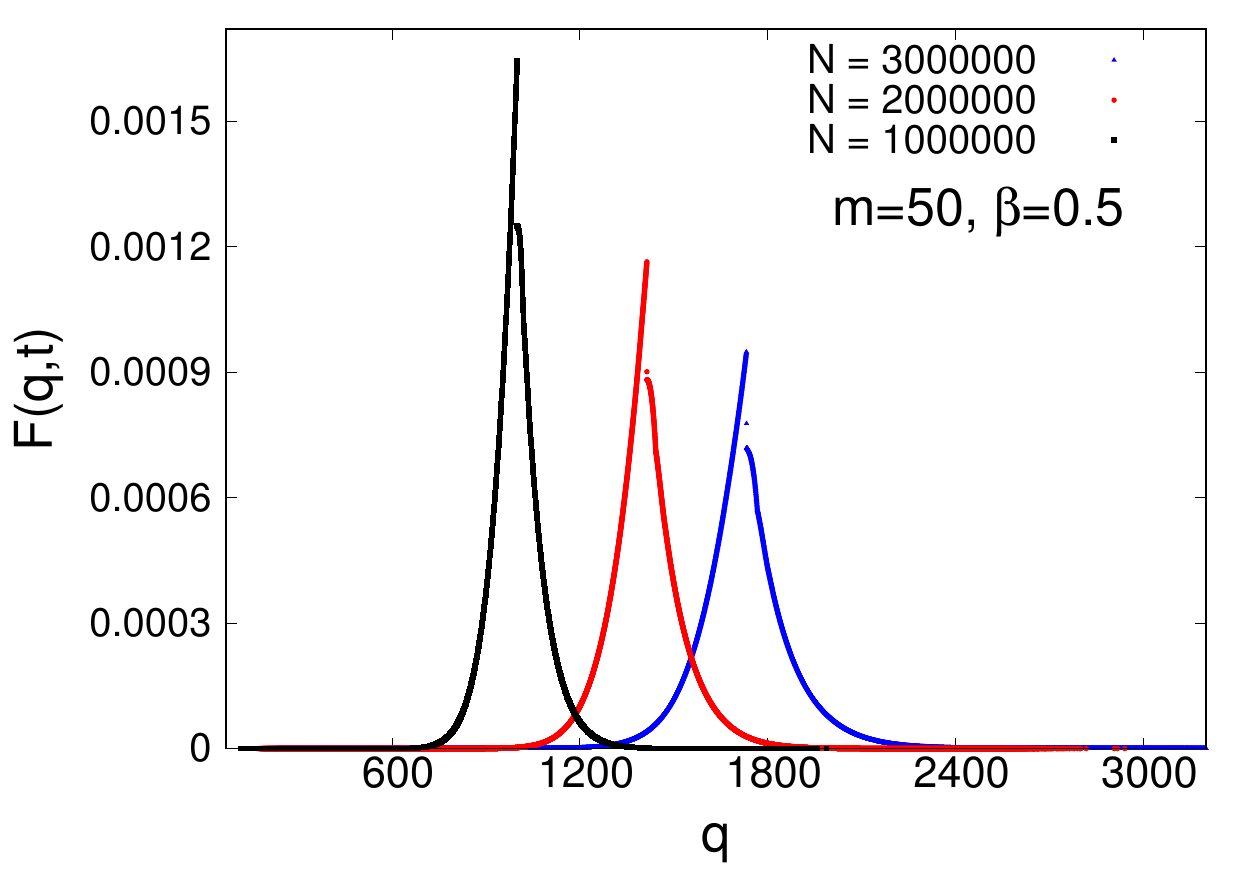}
\label{fig:6e}
}
\subfloat[]
{
\includegraphics[height=2.4 cm, width=4.0 cm, clip=true]
{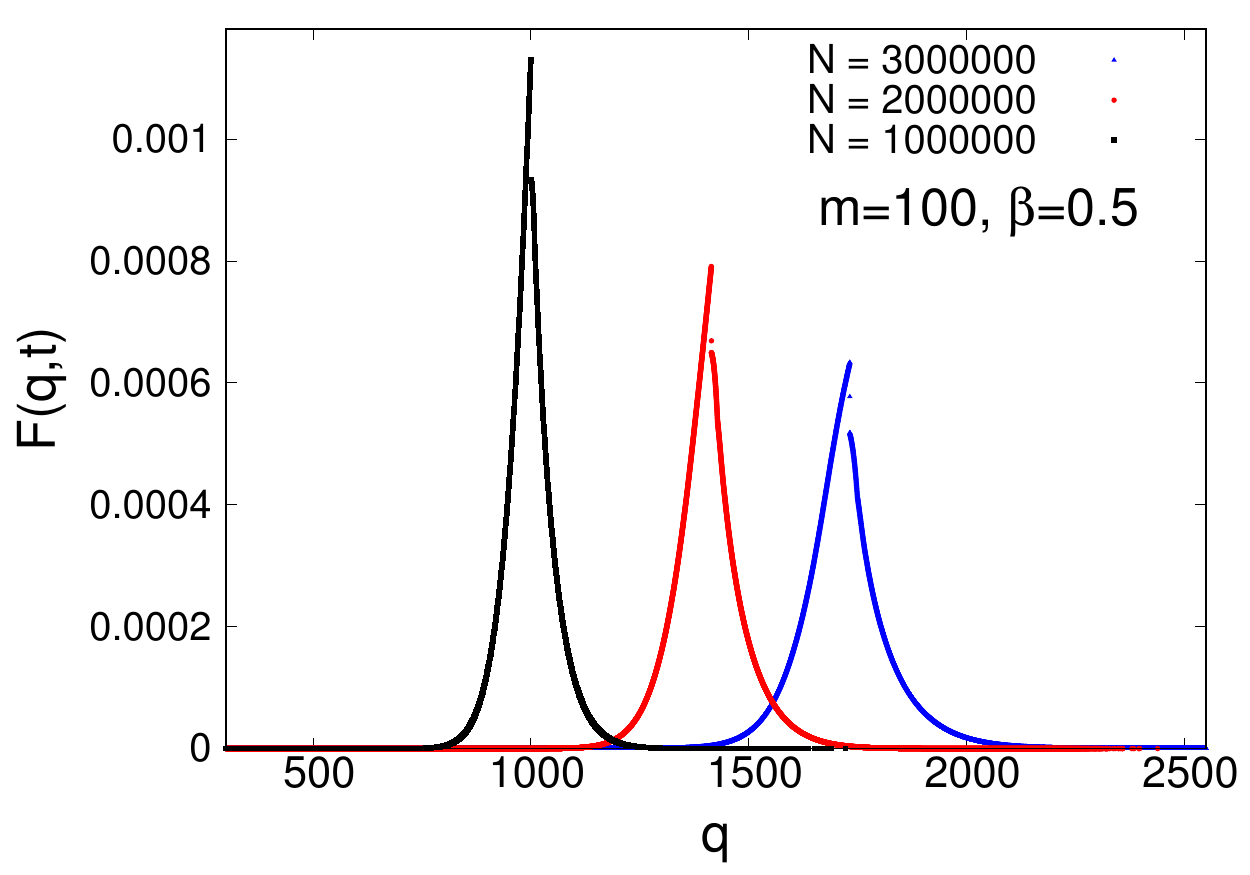}
\label{fig:6f}
}

\caption{
Plots of the generalized degree distribution $F(q,t)$ vs $q$ of the BA model for (a) $m=1$,(b)$m=3$, 
(c) $m=10$, (d) $m=15$, (e) $m=50$ and (f) $m=100$. These plots clearly show that for smaller
$m$ the difference between the two models is quite sharper. However, for larger $m$ they get closer.
} 

\label{fig:6abcdef}
\end{figure}

\section{Numerical Proof of Dynamic Scaling}

To prove that the generalized degree distribution of the MDA network exhibits dynamic scaling, we 
divide the abscissa and ordinate of the data of Figs. (\ref{fig:5abcdef}) 
by the corresponding $t^{-\beta}$ and $t^{\beta}$ respectively. Note that here $t$ is taken as the respective
time when the snapshot of the growing MDA network is taken. For instance, 
we took three snapshots at three different times $t_1, t_2, t_3$ 
of network sizes 1000k, 2000k and 3000k respectively while the
network has been growing with time following the MDA rule. 
This is akin to plotting $F(q,N)/N^{-\beta}$ versus
$\frac{q}{N^{\beta}}$ since $t\sim N$. 
In each plot, we find that all three distinct curves for 
different sizes of Figs. (\ref{fig:4abcdef}) collapse onto a single universal curve 
as shown in Figs. (\ref{fig:5abcdef}). The universal curve is
essentially the scaling function $\Phi(\frac{q}{t^\beta})$. We find that the quality of the 
data-collapse for $m=1$
to $m=11$ is slightly compromised owing to the wild fluctuation of the IHM as
reported in Ref. \cite{ref.hassan_liana}. However, for $m > 12$, 
the data-collapse is great due to the fact that from this value 
the mean of the IHM starts to have increasingly better meaning. 
Thus, dynamic scaling in networks grown using the
mediation-driven attachment model has been confirmed.

\begin{figure}

\centering

\subfloat[]
{
\includegraphics[height=2.4 cm, width=4.0 cm, clip=true]
{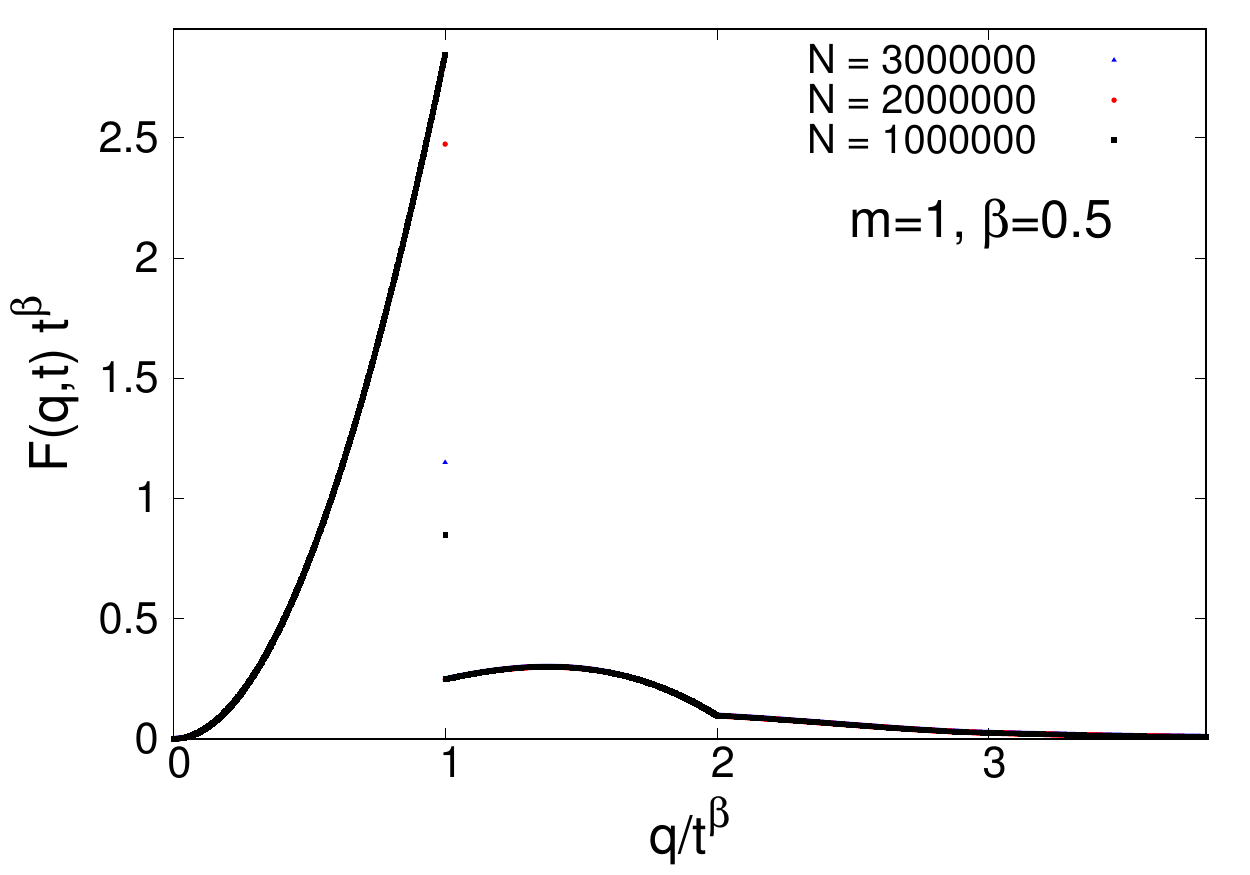}
\label{fig:7a}
}
\subfloat[]
{
\includegraphics[height=2.4 cm, width=4.0 cm, clip=true]
{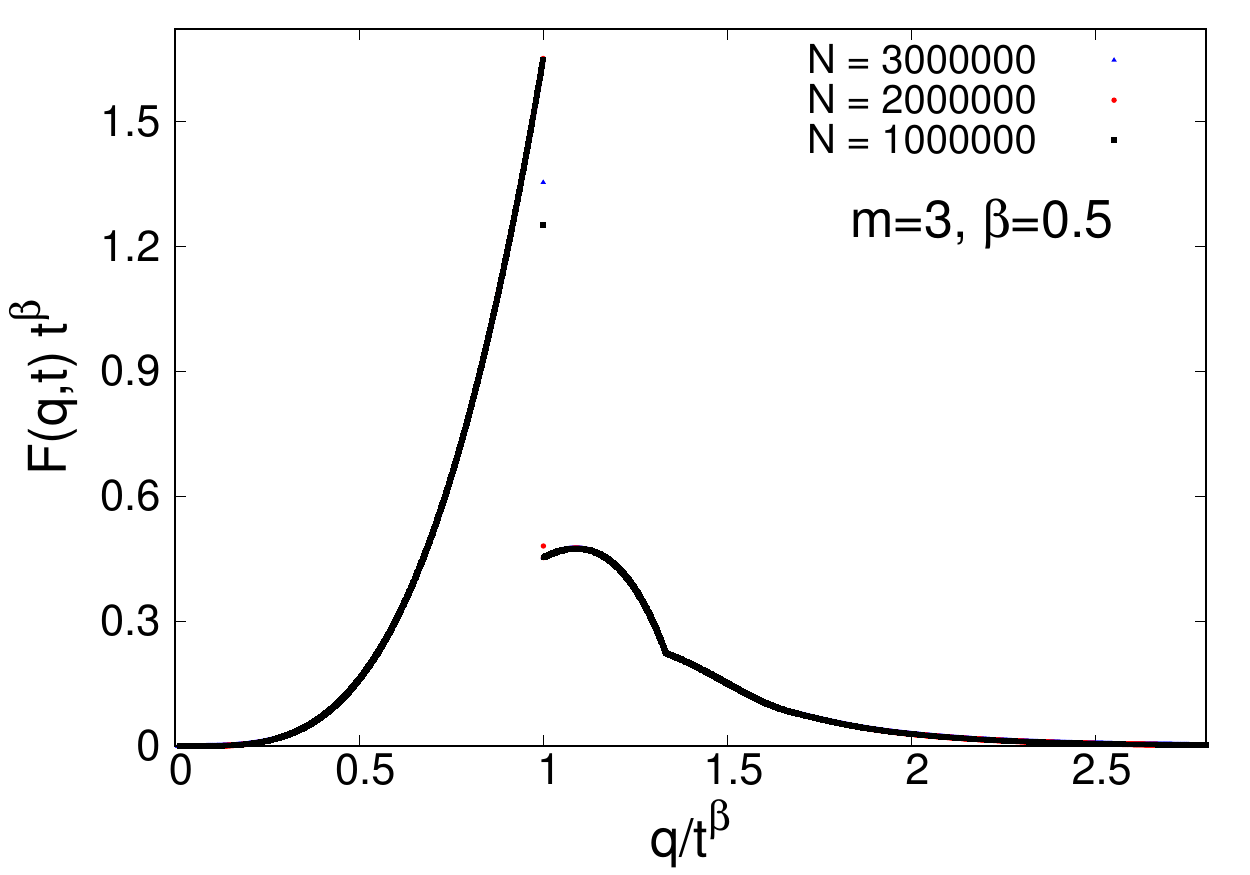}
\label{fig:7b}
}

\subfloat[]
{
\includegraphics[height=2.4 cm, width=4.0 cm, clip=true]
{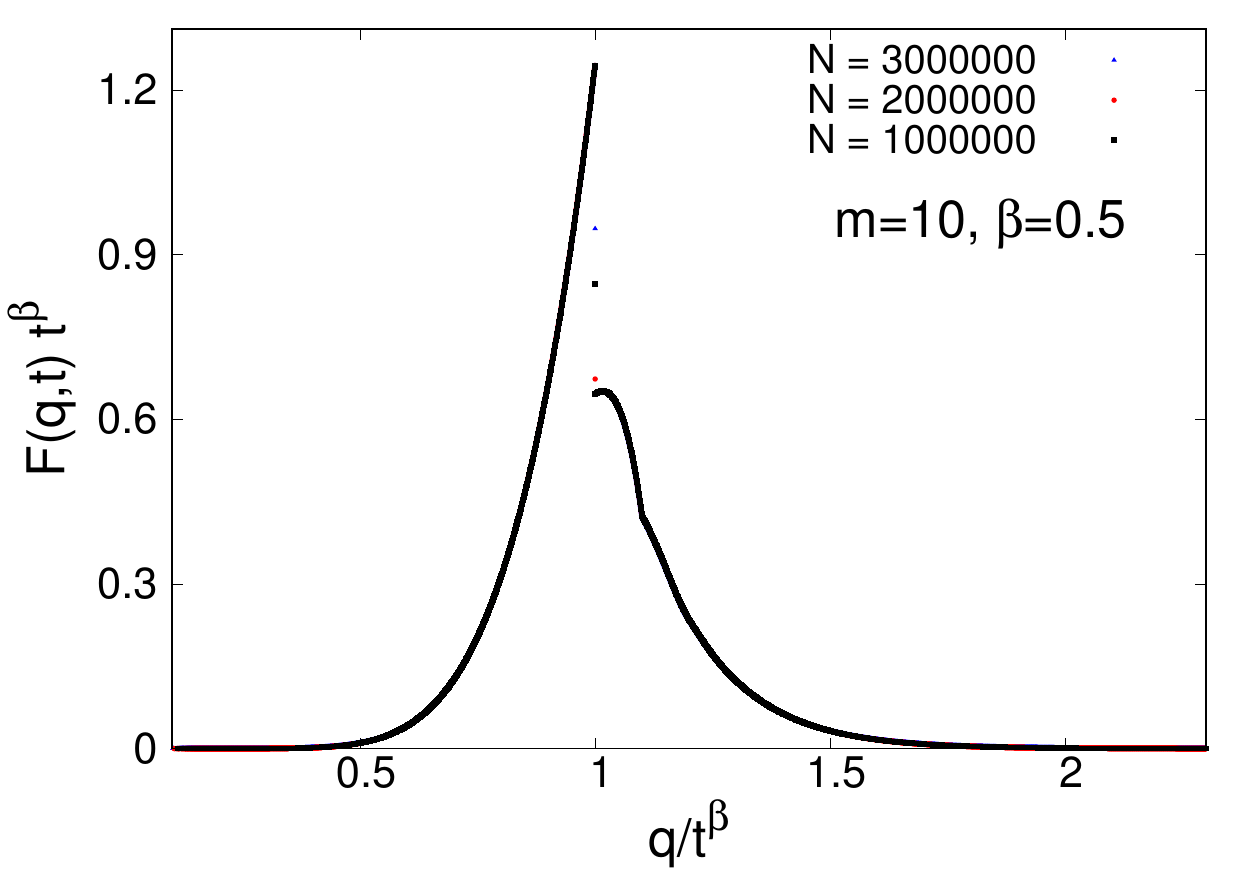}
\label{fig:7c}
}
\subfloat[]
{
\includegraphics[height=2.4 cm, width=4.0 cm, clip=true]
{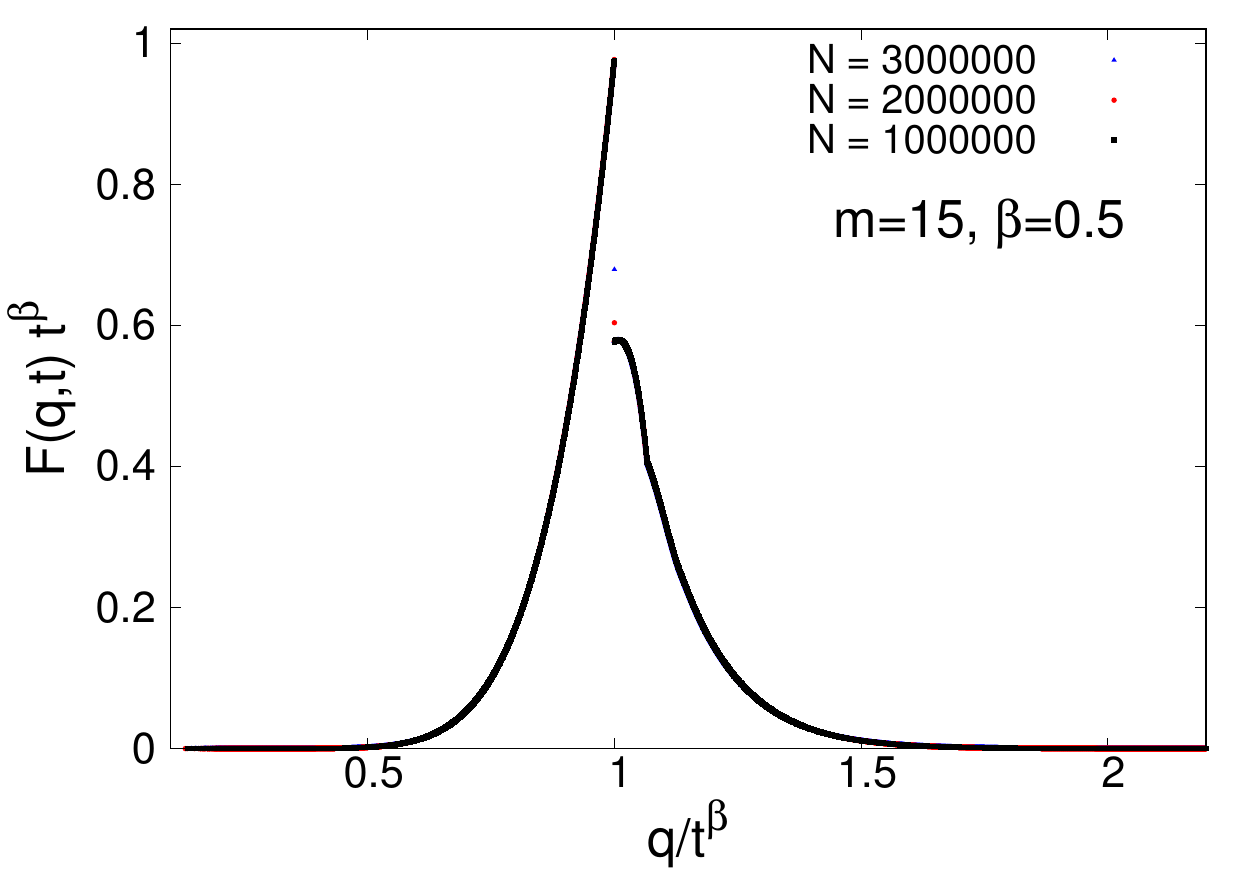}
\label{fig:7d}
}

\subfloat[]
{
\includegraphics[height=2.4 cm, width=4.0 cm, clip=true]
{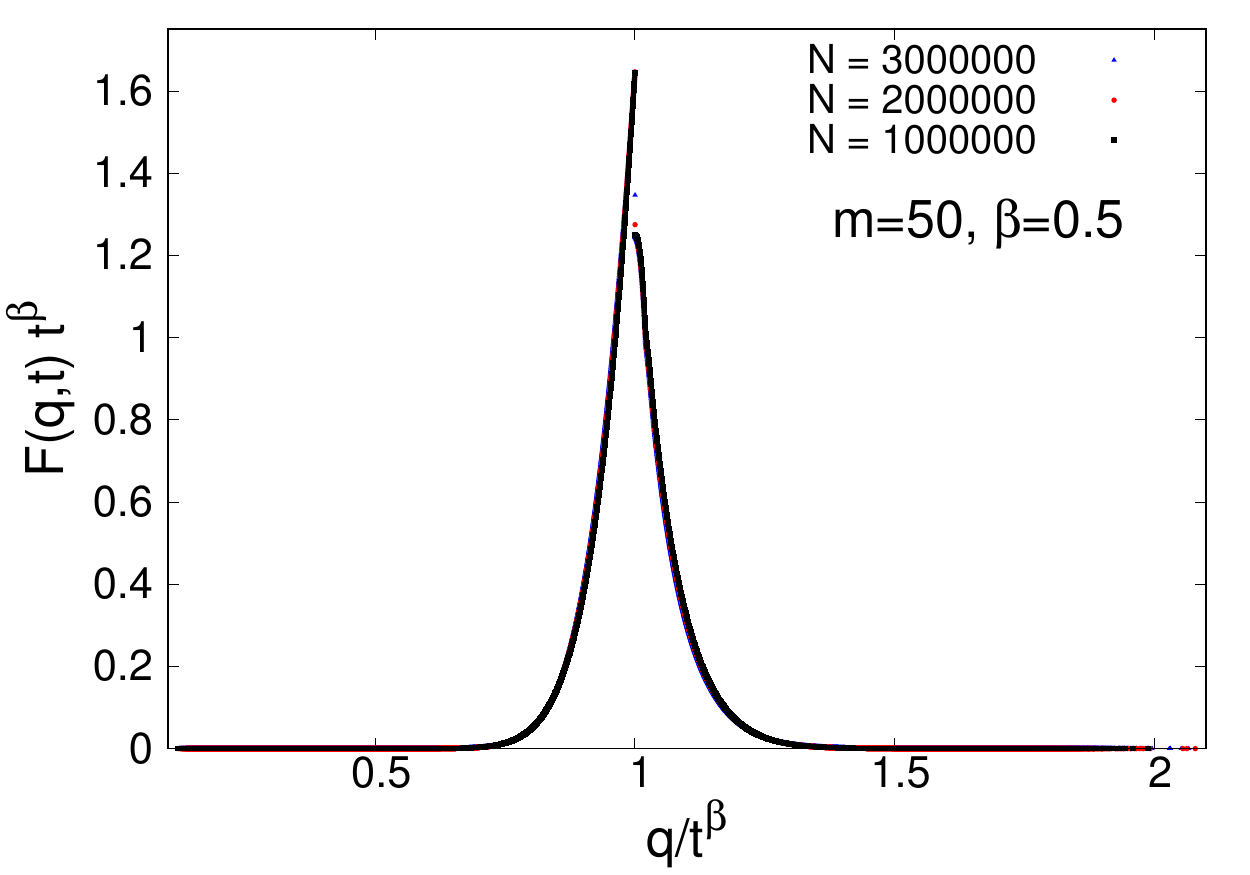}
\label{fig:7e}
}
\subfloat[]
{
\includegraphics[height=2.4 cm, width=4.0 cm, clip=true]
{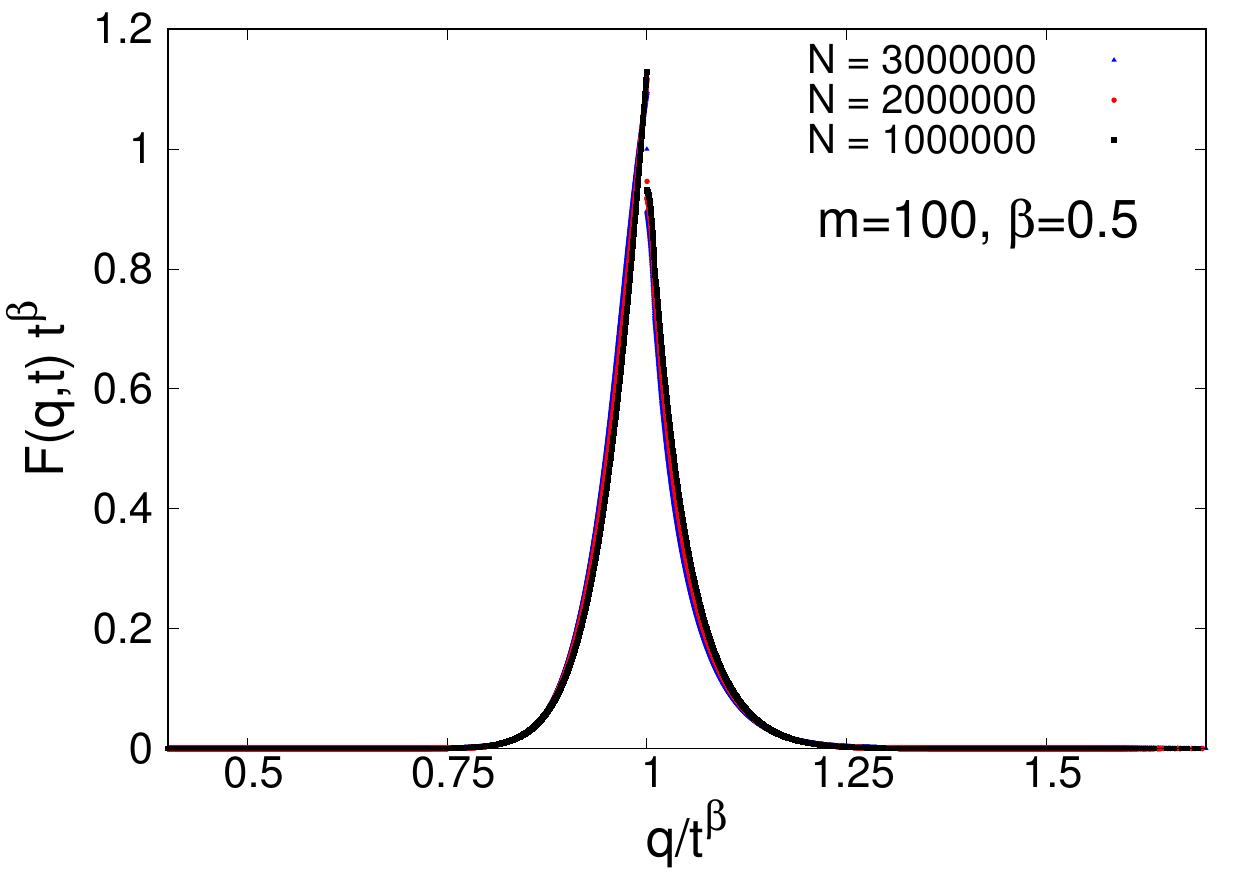}
\label{fig:7f}
}
\caption{We use the same data as in Fig. (\ref{fig:6abcdef}) for the BA model 
and plot $F(q,t)t^{1/2}$ versus
$qt^{-1/2}$  and find that for each value of $m$ the distinct plots in Fig. (\ref{fig:6abcdef}) collapse superbly
proving that BA networks grow following dynamic scaling.
} 

\label{fig:7abcdef}
\end{figure}

To see how our MDA model differs from the BA model, we show the plots of $F(q,t)$
versus $q$ in Figs. (\ref{fig:6abcdef}) and the plots for the corresponding data collapse
in Figs. (\ref{fig:7abcdef}). In particular, we find that the resulting scaling curve of the MDA model 
for $m=1$ is significantly different
from that of the BA model (see the Figs. (\ref{fig:4a}) and (\ref{fig:6a})).
 The generalized degree distribution curve of the MDA model for
$m=1$ has a flat horizontal line whose height and extent $q_c$ depend on the network size which reflects
the fact that the rich and poor gap is quite strong.  The higher the height the shorter the 
extent up to which the line exists.  Moreover, for 
small $m$ but $m>1$, the difference between scaling curve of the two models still persists but the 
gap between rich and poor increasingly decreases. In the case of small but $m>1$,
the scaling curve starts from zero, rises with convex curvature to a maximum and then drops 
suddenly with a gap
which is common to both the models. However,  for large $m$, the scaling curve looks almost the same. It implies
that $m$ really matters in determining the nature of the network. Recently, we have studied explosive
percolation in the BA network for different $m$ and found interesting results \cite{ref.hassan_digonto}.

\begin{figure}

\centering

\subfloat[]
{
\includegraphics[height=2.4 cm, width=4.0 cm, clip=true]
{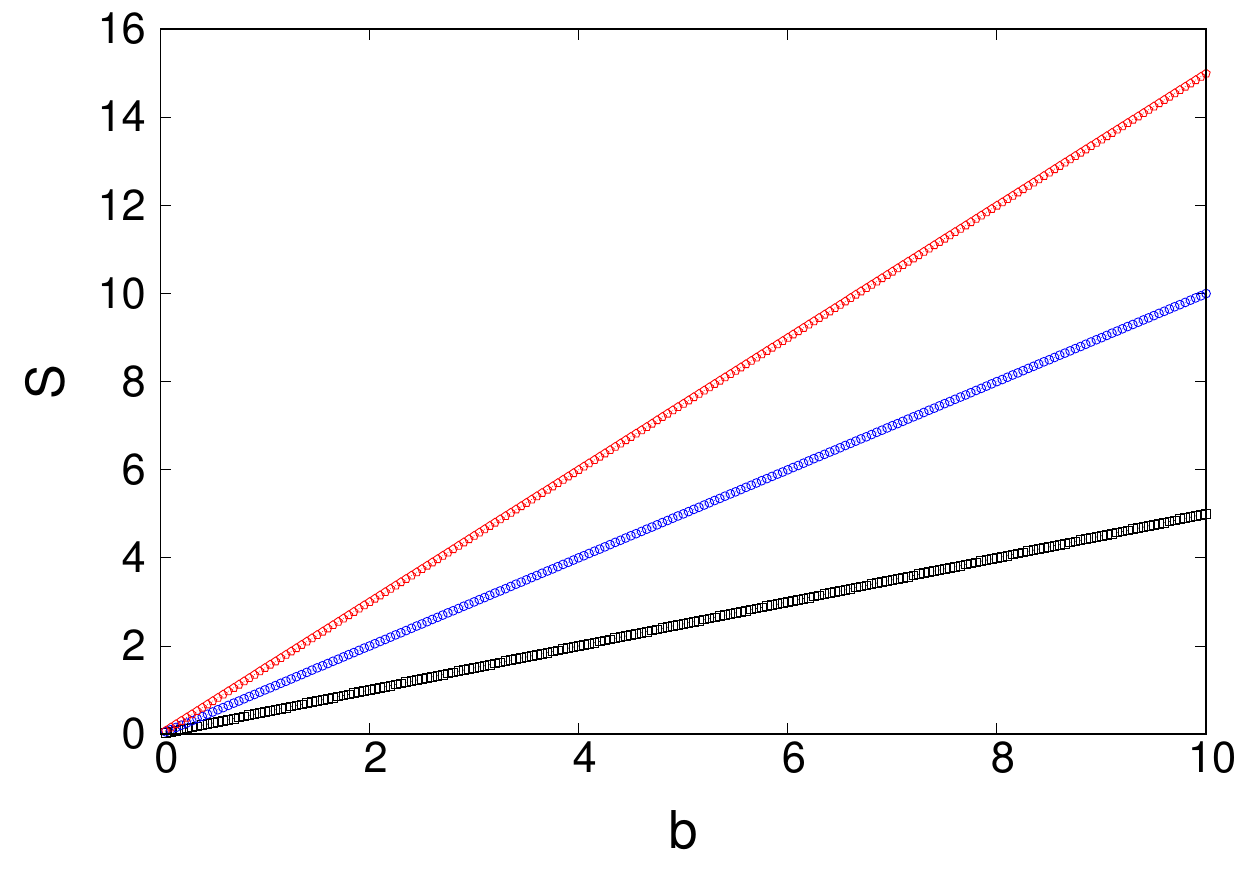}
\label{fig:8a}
}
\subfloat[]
{
\includegraphics[height=2.4 cm, width=4.0 cm, clip=true]
{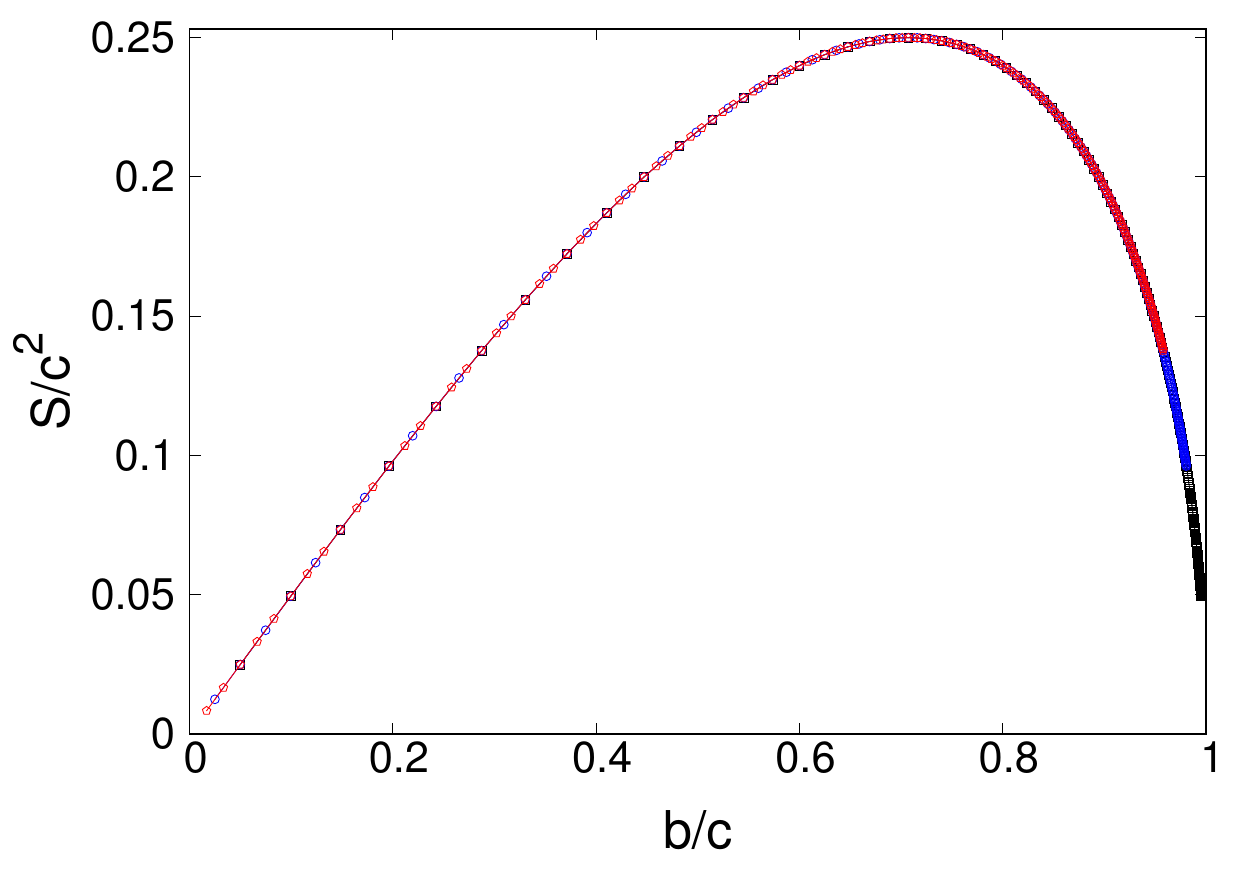}
\label{fig:8b}
}
\caption{(a) Areas $S$ of three right triangles $a$, whose adjacent sides are $1, 2$ and $3$, are plotted
as a function of their opposite sides $b$ that result in three distinct straight lines with slopes equal to
$0.5, 1.0$ and $1.5$ respectively. (b) We now measure the same areas $S$ in units of the square of their respective 
hypotenuse $c$ and the opposite side $b$ in unit of $c$ and plot them again. The resulting graph is equivalent to plotting $S/c^2$ versus $b/c$ and find that all the distinct plots of (a) collapse into a single universal curve.
}
\label{fig:8ab}
\end{figure}

\section{Significance of Dynamic Scaling and Data-collapse}

What can we conclude from a system that exhibits dynamic scaling?
If we know that a  time developing system exhibits dynamic scaling, it means that it is self-similar. 
By the term self-similarity, we mean that it is similar with itself at different times. Let us
briefly explain what we mean by {\it similarity}. Note that the same system at different times
is similar if the numerical values of various dimensional governing parameters are different. However, the numerical values of the 
corresponding dimensionless quantities coincide for a given value of dimensionless governing parameter. 
This idea of similarity can in fact be thought of as an extension of the criterion for geometric similarity. For instance, two triangles are 
said to be similar even if the numerical values of their sides are different but the corresponding dimensionless quantities such as their angles are the same. Consider that we have three triangles
of different sizes as shown in Fig. (\ref{fig:8a}) and we measure their area $S$ 
as a function of their height (opposite) side $b$  
keeping their base (adjacent) $a$ the same, the resulting plot will be a set of distinct straight lines with 
slopes equal to their respective $a/2$ since we know that $S=1/2 ab$.  
However, we find that all the distinct straight lines with different slopes collapse onto a single universal curve 
if we plot the dimensionless quantity $S/c^2$ instead of $S$ as a function of dimensionless quantity $b/c$ instead of $b$ as shown in Fig. (\ref{fig:8b}). It implies that for a given value of $b/c$ the numerical 
value of $S/c^2$ is the same regardless of the size of the triangle. Thus, data-collapse is indeed a
litmus test for similarity. However, if the same system at different times is similar to itself is 
said to be self-similar. In the same way extending the idea of data collapse to the BA or MDA networks 
 of different sizes is said to be similar.

\section{Conclusion}

In the MDA and BA networks, the generalized degree distribution function $F(q,t)$ exhibits dynamic scaling
$F(q,t \rightarrow \infty) \sim t^{-\beta} \Phi (q/t^\beta)$, where $\Phi (x)$ is the
scaling function. However, for the BA model $\beta=1/2$ regardless of the value of $m$ and for the
MDA model it depends on $m$ such a way that $1/\beta$ starts from $1$ and reaches asymptotically to the value
$2$ which coincides with that of the BA model. To prove the dynamic scaling we took a series of snapshots 
for each $m$. Then we extracted data from those snapshots and plotted generalized
degree distribution $F(q, N)$ as a function of $q$ for
different network sizes $N$, where $N \sim t$. The curves collapse onto a single universal curve
if $N^\beta F(q, N)$ is plotted against $q/N^\beta$. This implies that the
resulting networks of different sizes (at different times) are similar. 
 Finally, we found that the exponents of the MDA networks for different
values of $m$ are different. This is in sharp contrast
to what we find in the BA model where its exponent $\beta$ is independent of $m$.

\end{document}